\documentclass[journal,twoside]{IEEEtran}
\usepackage{cite}
\usepackage{graphicx}
\usepackage{comment}
\usepackage{balance}
\usepackage{acronym}
\usepackage{amsmath}
\usepackage{tabularx}
\usepackage[usenames,dvipsnames]{xcolor}
\usepackage{booktabs}
\usepackage{amssymb}
\usepackage{multirow}
\usepackage{subcaption}
\usepackage{hyperref}

\hypersetup{
 bookmarksopen=true,
 bookmarksnumbered=true,
 colorlinks=false,
 citecolor=blue,
 linkcolor=red,
 linktocpage=true,
 linkbordercolor={1 1 1},
 hypertexnames=false,
 plainpages=true,
 pdfsubject={},
 pdfkeywords={},
 pdftitle={},
 pdfauthor={}
}

\title{Explainable DNN-based Beamformer with Postfilter}

\author{Adi Cohen, Daniel Wong, Jung-Suk Lee and~Sharon Gannot,~\IEEEmembership{Fellow,~IEEE}
\thanks{Adi Cohen and Sharon Gannot are with the Faculty of Engineering, Bar-Ilan University, Ramat-Gan, Israel.
 e-mail: \texttt{Adi.Cohen5@biu.ac.il}; \texttt{sharon.gannot@biu.ac.il}}

\thanks{Daniel Wong and Jung-Suk Lee are with Meta Reality Labs, WA, USA. e-mail: \texttt{ddewong@meta.com};  \texttt{jungsuklee@meta.com}}

\thanks{This project has received funding from the European Union’s
Horizon 2020 Research and Innovation Programme, Grant Agreement
No. 871245, and from Meta Reality Labs.}
}

\acrodef{STFT}{Short-Time Fourier Transform }
\acrodef{ISTFT}{Inverse Short-Time Fourier Transform}
\acrodef{BSS}{Blind Source Separation}
\acrodef{DOA}{Direction of Arrival}
\acrodef{DC}{Deep Clustering}
\acrodef{DPRNN}{dual-path recurrent neural network}
\acrodef{TF}{Time-Frequency}
\acrodef{TCN}{Temporal Convolutional Network}
\acrodef{ATF}{Acoustic Transfer Function}
\acrodef{RTF}{Relative Transfer Function}
\acrodef{SISDR}{Scale-Invariant Signal-to-Distortion Ratio}
\acrodef{MSE}{Mean Square Error}
\acrodef{MAE}{Mean Absolute Error}
\acrodef{DFT}{Discrete Fourier Transform }
\acrodef{SIR}{signal-to-interference ratio}
\acrodef{OVA}{overlap-and-add}
\acrodef{SDR}{signal-to-distortion ratio}
\acrodef{BLSTM}{Bidirectional Long Short-Term Memory}
\acrodef{SOTA}{state-of-the-art}
\acrodef{RI}{Real-Imaginary}
\acrodef{RIR}{Room Impulse Response}
\acrodef{SNR}{signal-to-noise ratio}
\acrodef{RNN}{Recurrent Neural Networks}
\acrodef{SE}{Speech Enhancement}
\acrodef{SS}{Speech Separation}
\acrodef{MVDR}{Minimum Variance Distortionless Response}
\acrodef{MPDR}{Minimum Power Distortionless Response}
\acrodef{GEV}{Generalized Eigenvalue}
\acrodef{MWF}{Multichannel Wiener Filter}
\acrodef{SDW-MWF}{Speech Distortion Weighted Multichannel Wiener Filter}
\acrodef{MIMO}{Multi Input Multi Output}
\acrodef{LCMV}{Linearly Constrained Minimum Variance}
\acrodef{DNN}{Deep Neural Network}
\acrodef{PSD}{Power Spectral Density}
\acrodef{STOI}{Short Term Objective Intelligibility}
\acrodef{ESTOI}{Extended Short Term Objective Intelligibility}
\acrodef{PESQ}{Perceptual Evaluation of Speech Quality}
\acrodef{PF}{Post-Filter}
\acrodef{NR}{Noise Reduction}
\acrodef{AR}{Auto Regressive} 
\acrodef{PCA}{Principal Component Analysis}
\acrodef{GEVD}{Generalized Eigenvalue Decomposition}
\acrodef{EVD}{Eigenvalue Decomposition}
\acrodef{GSVD}{Generalized Singular Value Decomposition}
\acrodef{EM}{Expectation-Maximization}

\begin{document}

\markboth{IEEE/ACM Transactions on Audio, Speech, and Language Processing,~Vol.~x, No.~y, Jul.~2024}
{Cohen \MakeLowercase{\textit{et al.}}: Explainable DNN-based Beamformer with Postfilter}
\maketitle

\begin{abstract}
This paper introduces an explainable DNN-based beamformer with a postfilter (ExNet-BF+PF) for multichannel signal processing. Our approach combines the U-Net network with a beamformer structure to address this problem. The method involves a two-stage processing pipeline. In the first stage, time-invariant weights are applied to construct a multichannel spatial filter, namely a beamformer. In the second stage, a time-varying single-channel post-filter is applied at the beamformer output. Additionally, we incorporate an attention mechanism inspired by its successful application in noisy and reverberant environments to improve speech enhancement further.

Furthermore, our study fills a gap in the existing literature by conducting a thorough spatial analysis of the network's performance. Specifically, we examine how the network utilizes spatial information during processing. This analysis yields valuable insights into the network's functionality, thereby enhancing our understanding of its overall performance.

Experimental results demonstrate that our approach is not only straightforward to train but also yields superior results, obviating the necessity for prior knowledge of the speaker's activity. 
Audio samples are available on our project page.\footnote{\url{https://exnet-bf-pf.github.io/}} The code will also be available.\footnote{The ExNet-BF+PF code 
 was created by Bar-Ilan University and will be released on \url{https://github.com/AdiCohen501/ExNet-BF-PF}}
\end{abstract}


\section{Introduction}
Multi-microphone noise reduction is one of the most extensively researched topics in audio signal processing. Traditionally, spatial filters, often referred to as beamformers, are designed to meet specific optimization criteria~\cite{gannot2017consolidated}. Various beamforming design criteria were proposed, such as \ac{MVDR}~\cite{affes1997signal,MVDR_art}, \ac{GEV}~\cite{GEV}, \ac{GSVD}~\cite{doclo2002gsvd}, \ac{MWF}~\cite{MWF}, \ac{SDW-MWF}~\cite{Doclo05b}, and \ac{LCMV}~\cite{markovich2009multichannel}. Several works~\cite{zelinski1988microphone,meyer1997multi,lefkimmiatis2006optimum,marro1998analysis,schwartz2017multispeaker,balan2002microphone,cohen2003integrated,gannot2004speech} proposed to decompose the beamformer into spatial and spectral components. 

These beamforming techniques require robust control mechanisms to estimate their components accurately, a challenge addressed by several studies~\cite{laufer2018source,laufer2020global,laufer2021audio,higuchi2017online}.


The advent of deep learning has led researchers to investigate new methods for enhancing spatial filters. This field can be categorized into distinct approaches.

The first category involves beamformers with \ac{DNN} control. Studies in this category leverage neural networks to develop advanced mechanisms for controlling traditional beamformers. In~\cite{chazan2018lcmv,malek2017speaker}, the system uses signals from all microphones to detect speaker activity patterns, facilitating the estimation of the noise spatial \ac{PSD} matrix and the \ac{RTF} of the relevant sources. 
In~\cite{martin2020online}, an online multichannel speech enhancement system combines a recursive \ac{EM} algorithm for iteratively updating the model parameters to improve noise estimation accuracy with a \ac{DNN} for estimating the speech presence probability.
In~\cite{Schwartz2024}, a control mechanism is implemented as a multi-task \ac{DNN}, comprising speaker activity detector and \ac{DOA} estimation branches. 


A second category focuses on entirely \ac{DNN}-based techniques, aiming to leverage neural networks to enhance speech quality. For instance, the FaSNet~\cite{FaSNet} algorithm and its extension~\cite{IFaSNet} are time-domain multi-frame algorithms that combine multichannel signals based on cross-correlation features between the channels. The algorithms are optimized end-to-end and are applied in two paths. However, a drawback of these methods is that they do not preserve the general structure of the beamformer, making it challenging to interpret the operations of these networks.

The third category comprises neural network-based beamforming approaches, which maintain the beamformer criterion while learning some building blocks or weights by applying neural networks. Several contributions, such as~\cite{ADLMVDR, DeepWithMVDR, NICEbeamRLR}, propose an approach that combines a neural network with the conventional \ac{MVDR} beamformer. In these approaches, the network learns the \ac{MVDR} components, which are then utilized to calculate beamforming weights. While these methods focus on studying the individual components, alternative approaches enable the network to learn weights while preserving the beamforming architecture. For example, in~\cite{Walter2022}, a complex-valued neural network with an autoencoder architecture is presented, aiming to preserve phase information and spatial cues.
Furthermore, the work presented in~\cite{CausalUnet} integrates two elements: it first learns the weights using a U-Net architecture and then applies the beamforming operation with these weights. The U-Net comprises an encoder-decoder structure with skip connections, enabling the extraction of high-level and low-level features. This article serves as the starting point in developing our proposed architecture.

Although many works incorporate the beamforming operation, only a few present a spatial analysis to verify whether the network effectively utilizes spatial information. Notably, in~\cite{Walter2023}, the influence of different training targets on the network's spatial filtering behavior is analyzed by investigating their beampatterns and the network's performance using various performance measures. 


In our contribution, we combine a U-Net network with the beamformer structure. Our approach decomposes the network into spatial and spectral components through a two-stage processing method. The first stage, applied to the multichannel signal, maintains time-invariant weights, constituting the beamforming filters. Conversely, the second stage, applied at the output of the beamforming stage, is a single-channel post-filter employing time-varying weights. 
Attention-based methods have shown promise in enhancing speech signal quality in noisy and reverberant environments. Building on this potential, we incorporate an attention mechanism into the skip connections of the U-Net architecture, similar to the approach proposed in~\cite{Amazon2019}, to strengthen the connection between channels. Instead of directly concatenating the features from earlier layers at the same scale, we first multiply the skip-connection layers by an attention mask. 

Importantly, our paper provides a comprehensive analysis of the network's spatial performance, specifically examining how the network utilizes spatial information during processing.

The rest of the article is organized as follows. Sec.~\ref{sec:problem} formulates the problem, Sec.~\ref{sec:model} details the proposed method, Sec.~\ref{sec:setup} describes the experimental setup, Sec.~\ref{sec:study} presents the experimental study, and Sec.~\ref{sec:conclusion} concludes the paper.

\section{Problem Formulation} 
\label{sec:problem}
Let $y_m(t)$ be a noisy speech mixture recorded by the $m$-th microphone:
\begin{multline}
    y_m(t) = h_m(t)*d(t) + n_m(t) = x_m(t) + n_m(t),\\ {m=0,1,\ldots,M-1}\,
    \label{eq:mix_time}
\end{multline}
where $m$ represents the index of the microphone, with $M$ the total number of microphones, $d(t)$ is the clean signal, $h_m(t)$ represents the \ac{RIR} between the speaker and the $m$-th microphone, $x_m(t)$ denotes the speech signal as received by the $m$-th microphone, and $n_m(t)$ the corresponding noise signal.

In the \ac{STFT} domain and in vector form~\eqref{eq:mix_time} can be recast as, 
\begin{align}
\mathbf{y}(l,k) &= \mathbf{h}(l,k)d(l,k) + \mathbf{n}(l,k)\\
&= \mathbf{x}(l,k) + \mathbf{n}(l,k)
\label{eq:mix_STFT}
\end{align}
where $l \in \{0,\ldots, L-1\}$ and $k \in \{0,\ldots, K-1\}$ are the time-frame and the frequency-bin (TF) indexes, respectively, and $L$ and $K$ represent the total number of time-frames and frequency bands, respectively. $\mathbf{y}(l,k) = [y_0(l,k),\ldots,y_{M-1}(l,k)]^\top$ is an $M\times 1$ vector comprising all microphone signals. Correspondingly, $\mathbf{x}(l,k)$ and $\mathbf{n}(l,k)$ represent the signal and noise components, respectively, and $\mathbf{h}(l,k)$, the vector of \acp{ATF} from the source to all microphones. Note that we allow time variations of the \acp{ATF}.

The objective of the proposed method is to estimate the clean but reverberant speech signal captured by the reference microphone by applying a spatial filter:
\begin{equation}
\hat{x}(l,k) = \mathbf{w}^\textrm{H}(l,k)\mathbf{y}(l,k),
\label{eq:beamfomer}
\end{equation}
with $\hat{x}(l,k)$ representing the estimated clean speech and $\mathbf{w}(l,k)$ representing the time-varying complex-valued beamformer weights that the network should infer.

\section{Proposed Architecture}
\label{sec:model}
\begin{figure*}[ht]
\centering
    \includegraphics[width=17cm, height=4cm]{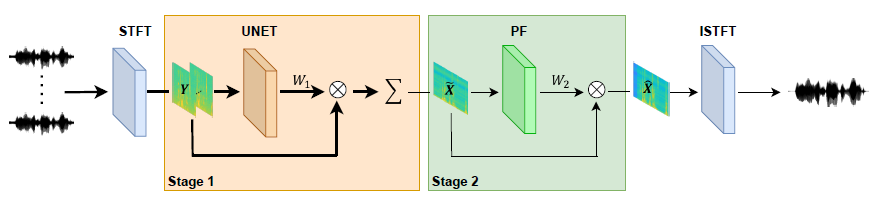}
    \caption{The ExNet-BF+PF, comprising two stages. In the first stage, a multichannel processor generates beamformer time-invariant weights. The second stage employs a single-channel post-filter with time-varying weights. In the figure, a thick arrow symbolizes multichannel processing, while a thin arrow symbolizes single-channel processing.}
    \label{fig:diagram}
\end{figure*}

This section presents the network architecture, decomposed into two stages, a multi-channel stage followed by a single-channel stage, as illustrated in Fig.~\ref{fig:diagram}. The inspiration for this decomposition comes from the well-known decomposition of the \ac{MWF} into a spatial filter, specifically the \ac{MVDR} beamformer, followed by a subsequent single-channel postfilter applied at the beamformer's output~\cite{MWF,balan2002microphone}. This architecture is termed ``explainable DNN-based beamformer with a postfilter (ExNet-BF+PF)''. 

The first stage of our architecture is a multichannel processor that implements a time-invariant beamformer with weights determined by a neural network. The second stage features a single-channel postfilter with time-varying weights, also implemented as a neural network.
\subsection{Multi-microphone Processing (Stage 1)}
In this stage, a U-Net architecture is employed to estimate the complex-valued weights of a filter-and-sum beamformer, represented as $\mathbf{w}_1(k)$. Subsequently, the output of the first stage $\Tilde{x}(l,k)$ is derived by the beamforming operation:
\begin{equation}
\Tilde{x}(l,k) = \mathbf{w}_1^\textrm{H}(k)\mathbf{y}(l,k),
\label{eq:beamformingOperationStage1}
\end{equation}
The input to this stage is $\mathbf{y}(l,k)$, structured into a 3-D tensor $\mathcal{Y}^c \in \mathbb{C}^{M \times K \times L}$, encompassing all frequencies and frames. The output of the beamformer stage, the enhanced speech in the \ac{STFT} domain, is given in tensor form as $\mathcal{\Tilde{X}} \in \mathbb{C}^{1 \times K \times L}$.

\subsubsection{U-Net Model}
The proposed U-Net model consists of an encoder and a decoder, with skip connections between them. Our architecture is based on~\cite{CausalUnet} but incorporates several modifications and improvements, particularly adjusting filter sizes across the frequency axis. 
The encoder comprises eight convolution layers, each of which is followed by batch normalization, dropout, and a `LeakyRelu' activation function. The decoder follows a similar architecture but with transpose-convolution layers instead of direct convolutions. The size of the filters, the kernel, and the stride in each layer are summarized in Table~\ref{table:Configuration_Table}.
\begin{table}[htbp]
\caption{The configuration of each Conv2d layer in the encoder.}
\begin{center}
\begin{tabular}{@{}ccccccc@{}}
\toprule
  Layer \# & \# of Filters & Kernel Size & Stride \\
  \midrule
1 & 32  & (6,3) & (2,2) \\ 
2 & 32  & (7,4) & (2,2) \\ 
3 & 64  & (7,5) & (2,2) \\ 
4 & 64  & (6,6) & (2,2) \\ 
5 & 96  & (6,6) & (2,2) \\ 
6 & 96  & (6,6) & (2,2) \\ 
7 & 128 & (2,2) & (2,2) \\ 
8 & 256 & (2,2) & (1,1) \\ 
\bottomrule
\end{tabular}
\end{center}
\label{table:Configuration_Table}
\end{table}
Following the U-Net, a linear layer is applied to the frequency dimension, followed by a `Tanh' activation function. 

Importantly, an additional average layer is implemented at the output of U-Net to enforce the first stage to become time-invariant. This results in constant $\mathbf{w}_1(k)$ weights (a complex-valued scalar per frequency bin and microphone).

The input to the proposed model is a 3-D tensor $\mathcal{Y} \in \mathbb{R}^{M \times 2K \times L}$, which is obtained by concatenating the real and imaginary components along the frequency axis. The corresponding output can be similarly written as $\mathcal{W}_1 \in \mathbb{R}^{M \times 2K \times 1}$.

Three comments regarding the implementation are in place:
a) Considering that the real and imaginary components of the beamformer weights may take negative values, we have chosen to employ the `Tanh' activation function.
b) After applying the \ac{ISTFT}, we aim to obtain a real-valued signal. To achieve this, it is necessary for both frequency 0 and frequency $\pi$ of the spectrogram to be real-valued. As the network structure does not enforce this condition, we have added a constraint at the end of the U-Net, following the time averaging step, to guarantee this.
c) Unlike the conventional \ac{MVDR} method, which independently operates per frequency, our approach utilizes the network to learn from all frequencies simultaneously, thereby leveraging their inherent connections.

\subsubsection{Attention Blocks as Skip Connections}
\begin{figure*}[ht]
\centering
    \includegraphics[width=15cm, height=8cm]{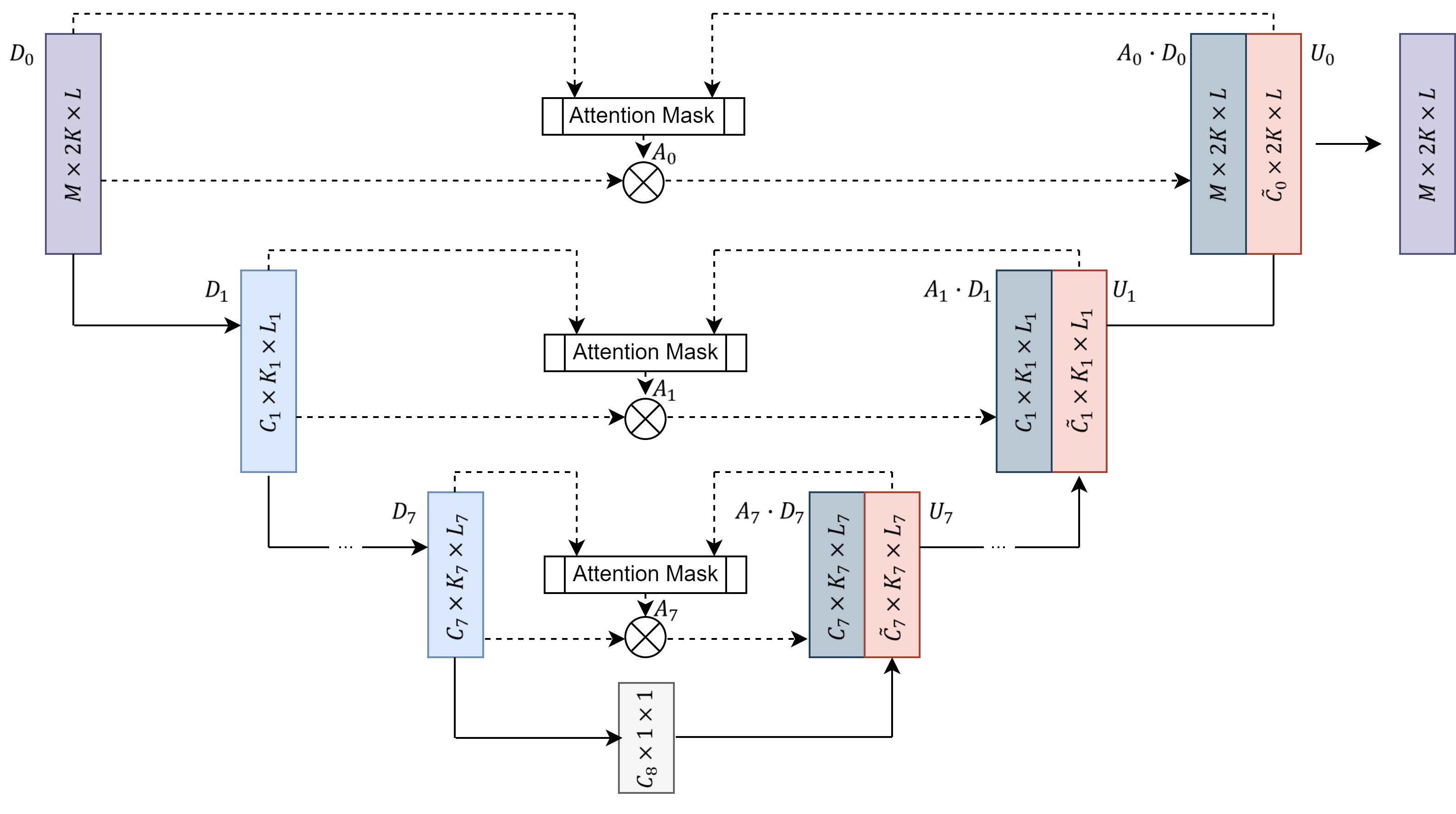}
    \caption{The proposed U-Net architecture. Encoder blocks ($D_i$) are colored in blue, decoder blocks ($U_i$) in red, and blocks subsequent to the attention mechanism ($A_i \cdot D_i$) are shaded in gray. Purple blocks indicate the input and output.}
    \label{fig:UNETFig}
\end{figure*}

Our architecture diverges from the traditional U-Net approach of directly concatenating features between the encoder and the decoder blocks (skip connections). Instead, we introduce attention gates to selectively identify relevant features from the encoder block, as shown in Fig.~\ref{fig:UNETFig}. This concept extends an idea introduced in~\cite{Amazon2019} and adapts it to the \ac{STFT} domain and the multichannel case. 
To implement the attention mechanism, we process each encoder block $D_i \in \mathbb{R}^{C_i \times K_i \times L_i}$ and its corresponding decoder block $U_i \in \mathbb{R}^{\Tilde{C}_i \times K_i \times L_i}$ at the same hierarchical level. These blocks are fed to an attention mask block that operates as follows:
\begin{itemize}
    \item The attention mask block comprises two separate 2-D convolution layers, one for $D_i$ and one for $U_i$. Each layer has a kernel size of (1,1) and $C_i/2$ filters.
    \item The outputs of these convolutional layers are summed and then passed through a `Sigmoid' activation function.
    \item The resulting output is processed by another 2-D convolutional layer with a kernel size of (1,1) and a single filter, followed by another `Sigmoid' activation function. This produces an attention mask $A_i\in \mathbb{R}^{1 \times K_i \times L_i}$.
\end{itemize}
The attention mask $A_i$ is subsequently multiplied element-wise with the encoder block $D_i$. The result of this multiplication is concatenated with the decoder block $U_i$. This combined output serves as the input to the next layer in the network. In the final layer, just before the dense layer, the attention mask is computed using the same procedure, with $D_0$ corresponding to the input of the U-Net model.

\subsection{Single-microphone Post-filtering (Stage 2)}
In this stage, a U-Net architecture is employed to estimate a complex mask, denoted as $w_2(l,k)$. Subsequently, the output of the second stage is a further enhanced speech signal denoted $\hat{x}(l,k)$, obtained by the following equation:
\begin{equation}
\hat{x}(l,k) = w_2^*(l,k)\Tilde{x}(l,k).
\label{eq:beamformingOpreationStage2}
\end{equation}
The input to this stage is the output of the first stage, and its output is the enhanced speech in the \ac{STFT} domain, denoted by $\hat{\mathcal{X}} \in \mathbb{R}^{1 \times K \times L}$. The corresponding time-domain signal, $\hat{x}(t)$, is obtained using the \ac{ISTFT} operation.
The second stage of the architecture closely resembles the first stage, with a few differences: (1) The input for the second stage is single-channel, differing from the multi-channel input of the first stage. (2) After the linear layer, the activation function is here a `Sigmoid'. (3) Unlike the first stage, no averaging was performed at the end of the U-Net in the second stage.

\subsection{Objectives}
In training the proposed model, the time-domain \ac{MAE} loss function is used, formulated as:
\begin{equation}
    \mathcal{L}_{\text{MAE}} = \textrm{mean}(|x_{\textrm{ref}}(t) - \hat{x}(t)|)
\end{equation}
where $x_{\textrm{ref}}(t)$ represents the target signal in the time domain as captured by the reference microphone, and $\hat{x}(t)$ denotes the estimated signal in the time domain. The mean operation calculates the average over the different times. 

A regularization term is incorporated into the loss function to improve the training procedure further. First, we apply the beamformer weights $\mathbf{w}_1(k)$, which were estimated using the noisy signals to the clean and reverberant speech signals, as captured by the microphones, $\mathbf{x}(l,k)$, instead of the noisy speech signals $\mathbf{y}(l,k)$. The output of this operation is denoted  $x_d(l,k)=\mathbf{w}_1^\textrm{H}(k)\mathbf{x}(l,k)$. Subsequently, we apply the \ac{ISTFT} to $x_d(l,k)$ to obtain the time-domain signal $x_d(t)$. The regularization term $\mathcal{L}_{\textrm{Reg}}$ is then computed as:
\begin{equation}
\mathcal{L}_{\textrm{Reg}} = \textrm{mean}(|x_{\textrm{ref}}(t) - x_d(t)|).
    \label{eq:loss_mae}
\end{equation}
This additional term encourages preserving a distortionless response towards the desired signal.

The final training loss is obtained as a weighted sum of the primary loss and the regularization term:  
\begin{equation}
    \mathcal{L} = \beta_{\textrm{MAE}}\cdot \mathcal{L}_{\textrm{MAE}} + \beta_{\textrm{Reg}} \cdot \mathcal{L}_{\textrm{Reg}}
\end{equation}
with $\beta_{\textrm{MAE}}+\beta_{\textrm{Reg}}=1$.

\section{Experimental setup}
\label{sec:setup}
This section details the experimental setup, covering the datasets employed for training and testing, algorithm settings, competing approaches, and evaluation measures. 

\subsection{Datasets}

For training the network, we constructed a dataset comprising noisy signals simulated in diverse environmental conditions. The dataset comprises 15,000 training samples, 5,000 validation samples, and 100 test samples for each environmental condition, ensuring a gender-balanced distribution of speakers.

\subsubsection{Database in a Non-reverberant Environment}
\begin{figure}[ht]
\centering
    \includegraphics[width=\linewidth]{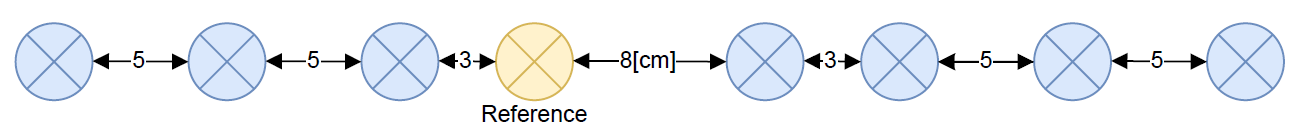}    
    \caption{The array configuration. The microphone inter-distances are given in centimeters (cm).}
    \label{fig:array_config}
\end{figure}

Each signal within this dataset consists of a target speaker, directional noise with a \ac{SNR} of 3~[dB], and additional spatially and spectrally white noise with an \ac{SNR} of 30~[dB]. This noise imitates sensor noise and is also useful as a regularization for the noise matrix inversion in the \ac{MVDR} construction, as will be explained later.
The target speaker is randomly selected from the LibriSpeech dataset~\cite{panayotov2015librispeech}. The directional noise is a colored noise modeled as an \ac{AR} process of order 1 with coefficient $-0.7$. Room dimensions, microphone array position, and the positions (radius $R$ and angle $\theta$ relative to the microphone array) of both the directional noise and the target speaker are randomly set in specified ranges, as detailed in Table~\ref{table:reverb_parameters}. The radius of the noise and the speaker are identical. We also ensured that the angle between the sources is at least $20^\circ$. The microphone array configuration is illustrated in Fig.~\ref{fig:array_config} with the array axis not parallel to the room walls. The duration of each signal in this dataset is set to 4 seconds, with the first half-second exclusively noise. 

\subsubsection{Dataset in a Reverberant Environment}
Similar to the non-reverberant dataset, signals were randomly drawn from the LibriSpeech dataset~\cite{panayotov2015librispeech}. However, each signal in this dataset was convoluted with a simulated \ac{RIR} using the \ac{RIR} generator tool~\cite{habets2006room}. The range of reverberation levels is specified in Table~\ref{table:reverb_parameters}. Room dimensions, microphone array position, and the directional noise and target speaker positions are determined as in the non-reverberant dataset.

\subsubsection{Noise Types}
Signals with different types of noise were generated for both reverberant and non-reverberant environments:

\begin{enumerate}
    \item Time-Varying Noise: The noise direction switched after 2~sec.
    \item Speaker Switching: The speaker and its direction switched after 2~sec.
    \item Babble Noise: Babble noise was generated by combining ten colored noise signals at various distances, maintaining an \ac{SNR} of 3~dB.
    \item Babble Voice: Babble voice was generated by combining ten speakers at different distances, maintaining an \ac{SNR} of 3~dB.
\end{enumerate}
For each environmental condition, the dataset comprised an additional 5,000 examples for each type of noise and an additional 100 test samples.

\begin{table*}[t]
\caption{Specification of the noisy and reverberant data.}
\label{table:noisy_data}
\centering
\resizebox{1.4\columnwidth}{!}{
\begin{tabular}{@{}lll@{}}
\toprule
Parameter & Symbol & Range/Value \\
\midrule
                    & $L_x$ & \emph{U}[6,9] \\
Room dimensions [m] & $L_y$ & \emph{U}[6,9] \\
                    & $L_z$ & $3$ \\ \midrule
Reverberation time [sec] & $T_{60}$ & \emph{U}[0.3,0.5] \\ \midrule
                               & $x$ & \emph{U}[2.5,${L_x}$-2.5] \\
Microphone Center Position [m] & $y$ & \emph{U}[0.5,${L_y}$-2.5] \\  
                               & $z$ & $1$ \\ \midrule
Tilt angle of the microphone array [$^\circ$] & $\phi$ & \emph{U}[-45,45] \\ \midrule
Speaker \& directional noise angle [$^\circ$] & $\theta$ & \emph{U}[0,180] (angle between the sources at least $20^\circ$) \\ \midrule
Speaker \& directional noise radius [m] & $R$ & \emph{U}[1.8,$\min(x-0.5,L_x-x-0.5,L_y-y-0.5,2.2)$] \\
\bottomrule
\end{tabular}
}
\label{table:reverb_parameters}
\end{table*}

\subsection{Algorithm Settings}
The speech and noise signals have a sample rate of 16~KHz. The \ac{STFT} frame size is set to 512 samples with $75\%$ overlap. Due to the symmetry of the \ac{DFT}, only the first half of the frequency bands are processed. The real and imaginary components of the spectrograms are concatenated. For the training procedure, we employed the Adam optimizer~\cite{kingma2014adam} with a learning rate of 1e-4 and a batch size of 16. The weights were initialized randomly, and the length of the signals was set to 4 seconds. 

\subsection{Competing Approaches}
\label{sec:CompetingApproaches}
%

Our assessment compares the proposed algorithm, ``ExNet-BF+PF," to the \ac{MVDR} beamformer. Various approaches can be employed to derive the \ac{MVDR} beamforming weights. For our study, the \ac{MVDR} weights $\mathbf{w}_{\textrm{mvdr}}(k)$ at frequency index $k$ are then given by:
\begin{equation}
    \mathbf{w}_{\textrm{mvdr}}(k) = \frac {\boldsymbol{\Phi}_{\mathbf{nn}}^{-1}(k)\tilde{\mathbf{h}}(k)}{\tilde{\mathbf{h}}^\textrm{H}(k)\boldsymbol{\Phi}_{\mathbf{nn}}^{-1}(k)\tilde{\mathbf{h}}(k)};\; \quad \mathbf{w}_{\textrm{mvdr}}(k) \in \mathbb{C}^M
\label{eq:MVDRsolution}
\end{equation}
where $(\cdot)^\textrm{H}$ is the Hermitian operator, $\boldsymbol{\Phi}_{\mathbf{nn}}$ the spatial covariance matrix of the noise, and $\tilde{\mathbf{h}}(k)\triangleq\frac{\mathbf{h}(k)}{h_{\textrm{ref}}(k)} \in \mathbb{C}^M$ indicates the vector of \acp{RTF}. Each vector component is defined as the ratio between an \ac{ATF} from the source to a microphone normalized by the reference microphone's respective \ac{ATF}.\footnote{While, in general, the \acp{ATF} and the corresponding \acp{RTF} may change over time, we omit here the time dependency, mainly due to our estimation procedure that estimate the signals' correlation matrices by time-averaging on all available frames, see \eqref{eq:noise_corr} and \eqref{eq:noisy_corr}.} The vector of \acp{RTF} constitutes 
the steering vector towards the target speech \cite{gannot2001signal}. 

Since the real values are unavailable, an estimate is required. The noise covariance matrix  $\boldsymbol{\Phi}_{\mathbf{nn}}$ is estimated from the data as follows:
\begin{align}
\hat{\boldsymbol{\Phi}}_{\mathbf{nn}}(k) =& \frac{1}{L_n}\sum_{l=0}^{L_n-1}\mathbf{y}(l,k)\mathbf{y}^\textrm{H}(l,k)\notag\\
=& \frac{1}{L_n}\sum_{l=0}^{L_n-1}\mathbf{n}(l,k)\mathbf{n}^\textrm{H}(l,k),
\label{eq:noise_corr}
\end{align}
where $L_n$ represents the number of noise-only time-frames at the beginning of the utterance, determined using prior knowledge provided in advance to the \ac{MVDR}. We stress that our network does not utilize this prior information.

In the same vein as the \ac{GEVD} procedure presented in~\cite{markovich2009multichannel}, the \ac{RTF} estimate is obtained by applying the \ac{EVD} procedure to the whitened correlation matrix of the noisy signal. 

The following procedure was applied. 
First, the spatial correlation matrix of the noisy signals is obtained by averaging over all time frames comprising the speech signal (in our case, from $L_n$, the last noise-only frame until the end of the utterance): 
\begin{equation}
    \hat{\boldsymbol{\Phi}}_{\mathbf{yy}}(k) = \frac{1}{L-L_n}\sum_{l=L_n}^{L-1}\mathbf{y}(l,k)\mathbf{y}^\textrm{H}(l,k).
    \label{eq:noisy_corr}
\end{equation}
The whitened measurements are given by:
\begin{equation}
    \mathbf{y}_{\textrm{w}}(l,k)=\hat{\boldsymbol{\Phi}}_{\mathbf{nn}}^{-1/2}(k)\mathbf{y}(l,k)
    \label{eq:white}
\end{equation}
with 
\begin{equation}
   \hat{\boldsymbol{\Phi}}_{\mathbf{nn}}^{-1/2}(k)= \mathbf{V}(k)\mathbf{D}^{-1/2}(k)\mathbf{V}^\textrm{H}(k),
\end{equation}
where $\mathbf{V}(k)$ and $\mathbf{D}(k)$ are the matrix of eigenvectors and the diagonal matrix of eigenvalues of the noise correlation matrix, respectively, and $(\cdot)^{-1/2}$ is the inverse square root of the matrix.
The spatial correlation matrix of the whitened signals is given by:
\begin{equation}
   \hat{\boldsymbol{\Phi}}_{\mathbf{y}_{\textrm{w}}\mathbf{y}_{\textrm{w}}}(k)=\hat{\boldsymbol{\Phi}}_{\mathbf{nn}}^{-1/2}(k)\hat{\boldsymbol{\Phi}}_{\mathbf{yy}}(k)\hat{\boldsymbol{\Phi}}_{\mathbf{nn}}^{-1/2\textrm{H}}(k),
   \label{eq:white_corr}
\end{equation}
with $(\cdot)^{-1/2\textrm{H}}$ is the Hermitian of the inverse square root of the matrix.

Let $\mathbf{f}(k)$ be the eigenvector corresponding to the maximum eigenvalue of the matrix $\hat{\boldsymbol{\Phi}}_{\mathbf{y}_{\textrm{w}}\mathbf{y}_{\textrm{w}}}(k)$. 
The \ac{RTF} is obtained by multiplying this eigenvector by the square root of the noise correlation matrix:
\begin{equation}
    \tilde{\mathbf{f}}(k) = \hat{\boldsymbol{\Phi}}_{\mathbf{nn}}^{1/2}(k)\mathbf{f}(k),
\end{equation}
where $(\cdot)^{1/2}$ denotes the square root of the noise correlation matrix, followed by a
normalization by the entry corresponding to the reference signal:
\begin{equation}
    \tilde{\mathbf{h}}(k)=\frac{\tilde{\mathbf{f}}(k)}{{\tilde{f}}_{\textrm{ref}}(k)}. 
\end{equation}

We have also implemented \ac{MPDR} beamformer as another baseline method. 
The \ac{MPDR} beamformer is given by:
\begin{equation}
    \mathbf{w}_{\textrm{mpdr}}(k) = \frac {\boldsymbol{\Phi}_{\mathbf{yy}}^{-1}(k)\tilde{\mathbf{h}}(k)}{\tilde{\mathbf{h}}^\textrm{H}(k)\boldsymbol{\Phi}_{\mathbf{yy}}^{-1}(k)\tilde{\mathbf{h}}(k)};\; \quad \mathbf{w}_{\textrm{mpdr}}(k) \in \mathbb{C}^M.
\label{eq:MPDRsolution}
\end{equation}
The \ac{MPDR} beamformer does not require the spatial correlation matrix of the noise and uses instead the spatial correlation matrix of the noisy signals. 
Hence, one may assume that prior information on noise-only frames is unnecessary. This is true for the conventional implementation of the \ac{MPDR} beamformer.   
However, our implementation utilizes the \ac{RTF} as a steering vector. Estimating the \ac{RTF} necessitates the noise spatial correlation matrix for the whitening operation, as shown in \eqref{eq:white}. 
To circumvent this requirement, we substituted the $\hat{\boldsymbol{\Phi}}_{\mathbf{nn}}(k)$ with a spatially-white noise correlation matrix $\sigma_n^2\mathbf{I}$. 

As the results of this variant were unsatisfactory, we will only report on the results for \ac{MVDR}, which relies on prior knowledge of noise-only segments, as explained earlier.

To guarantee a fair comparison between the \ac{MVDR} beamformer and our method, we also apply a \ac{PF} at the output of the \ac{MVDR} beamformer, referred to as ``MVDR+PF". We have selected the single-microphone log-spectral amplitude estimator as implemented in~\cite{PF}.  
Another baseline method is the second stage of the proposed scheme, namely the single-channel U-Net-based post-filter. We denote this baseline as ``SC". 

We also compare our method to the original ``CUNET'' method~\cite{CausalUnet}, which uses a causal network architecture. Considering that the CUNET model is a causal system, whereas ours is non-causal, we also configure the CUNET network in a non-causal mode, termed ``UNET". This involves running the network without zero-padding at the input, resulting in the modified network relying on future samples for its operation. Notably, due to the absence of an official code, we implemented our own version of this approach.

\subsection{Evaluation Measures}
To evaluate the performance of the proposed algorithm and compare it with the baseline approaches, we employ five evaluation measures: 1) \ac{SISDR}~\cite{SISDR}, 2) \ac{PESQ}~\cite{PESQ}, 3) \ac{STOI}~\cite{STOI}, 4) \ac{ESTOI}~\cite{ESTOI}, and 5) \ac{NR}, calculated as follows:

\begin{equation}
    \mathbf{NR} = 10 \log_{10}\left(\frac{\text{var}(\hat{x}_{\text{speech}}(t))}{\text{var}(\hat{x}_{\text{noise}}(t))}\right).
\label{eq:NR}
\end{equation}
Here, the subscript ${(\cdot)_{\text{noise}}}$ denotes the initial half-second of the signal containing only noise, while the subscript ${(\cdot)_{\text{speech}}}$ pertains to the subsequent 3.5 seconds of the signal.

\section{Experimental study}
\label{sec:study}
This section details the experimental study, encompassing the performance of the proposed algorithm and competing methods under non-reverberant and reverberant conditions. We also analyze the results and draw insights into the proposed method's performance, emphasizing its explainability. 

\subsection{Beampattern Analysis}
One of the main goals of our research is the explainability of the proposed \ac{DNN}-based scheme. Specifically, we aim to explore the spatial characteristics of the proposed beamformer and to examine whether it utilizes spatial information or solely focuses on spectral information. 

When delving into the analysis of beampatterns in reverberant environments, the reflection pattern of the \acp{RIR} relating the source and the microphones comes into play. To facilitate the understanding of these complex patterns, the methodology described in~\cite{Beampattern} introduces a simplified 1D representation of array responses. In this section, we apply this approach to comprehensively assess the spatial properties of our methods and compare them with the baseline methods outlined in Sec.~\ref{sec:CompetingApproaches} for both reverberant and non-reverberant environments. 

For this analysis, we position the speaker along a predefined semi-circular contour. Subsequently, we empirically generate the beampattern by moving a white-noise source around this perimeter and evaluating the array's response to this noise signal as a function of the \ac{DOA}. We stress that this evaluation considers reverberation, and the \ac{DOA} only serves as an attribute of the source location.

The computation of the beampower involves the following steps: a) Calculating the beamformer weights based on the chosen method. b) Using the \ac{RIR} generator~\cite{habets2006room}, we generate acoustic responses from the noise's positions to the microphone array, with $h_m(k,\theta)$ the corresponding \ac{ATF}. In non-reverberant environments, we employ \ac{ATF} with zero-order reflections, while in reverberant environments, we use \ac{ATF} with full-order reflections. Here, $\theta$ is measured relative to the array's axis, and $k$ indicates the $k-th$ frequency bin. c) Computation of the narrow-band beampattern as $B(k,\theta) = \mathbf{w}^\textrm{H}(k)\mathbf{h}(k,\theta)$. d) The wide-band beampower is finally obtained as $P(\theta) = \sum_k |B(k,\theta)|^2$. Typically, the wide-band beampower is visualized as a polar plot in dB scale, with the maximum value normalized to 0~dB.

\subsubsection{Dataset in a non-reverberant environment} 
\begin{figure}[t]
    \centering
    \begin{subfigure}{0.45\columnwidth}
        \includegraphics[width=\textwidth]{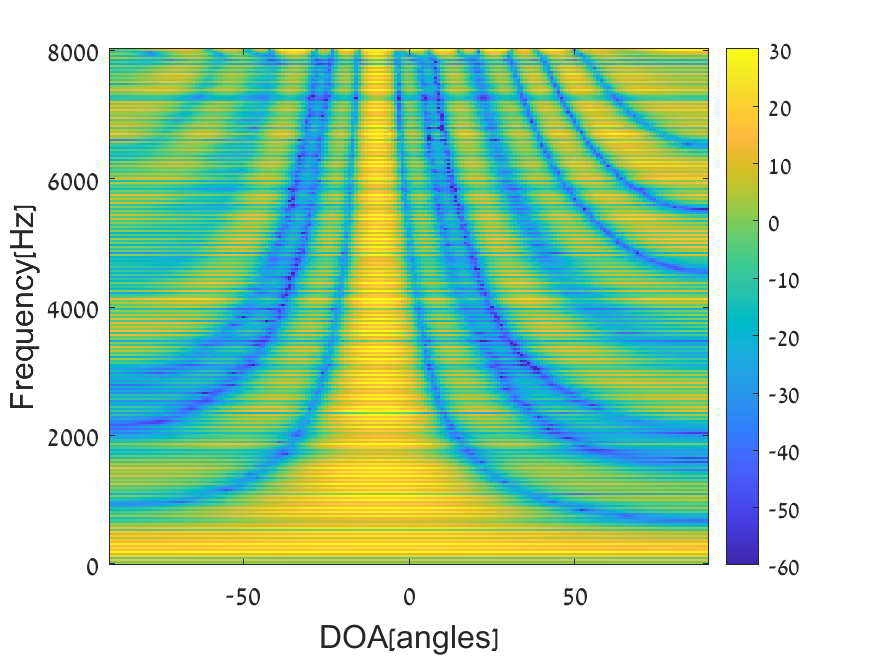}
        \caption{ExNet-BF+PF}
        \label{fig:FreqVSDoa_WithoutRevProposed}
    \end{subfigure}\hfill
    \begin{subfigure}{0.45\columnwidth}
        \includegraphics[width=\textwidth]{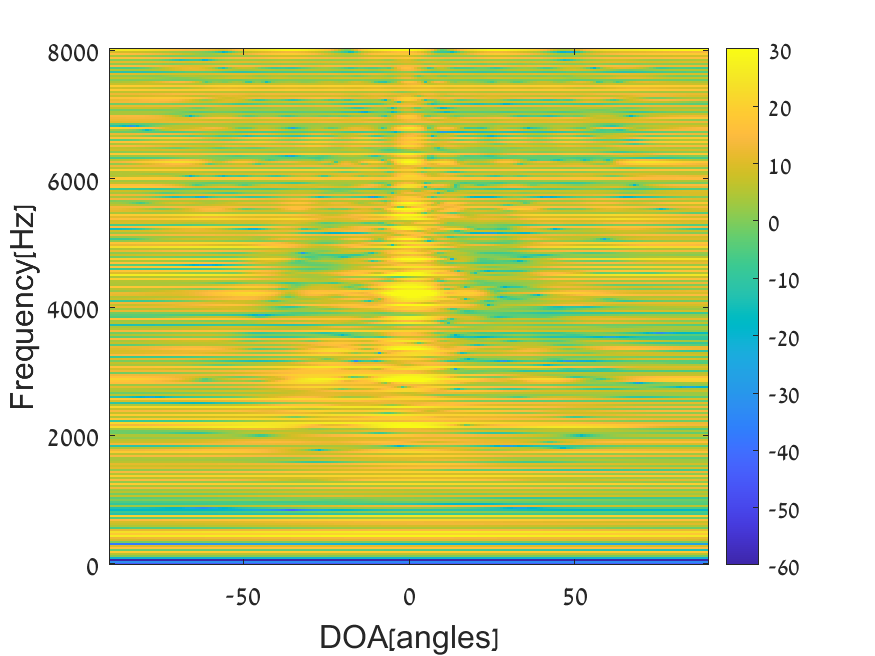}
        \caption{CUNET}
        \label{fig:FreqVSDoa_WithoutRevCUNET}
    \end{subfigure}    
    \begin{subfigure}{0.45\columnwidth}
        \includegraphics[width=\textwidth]{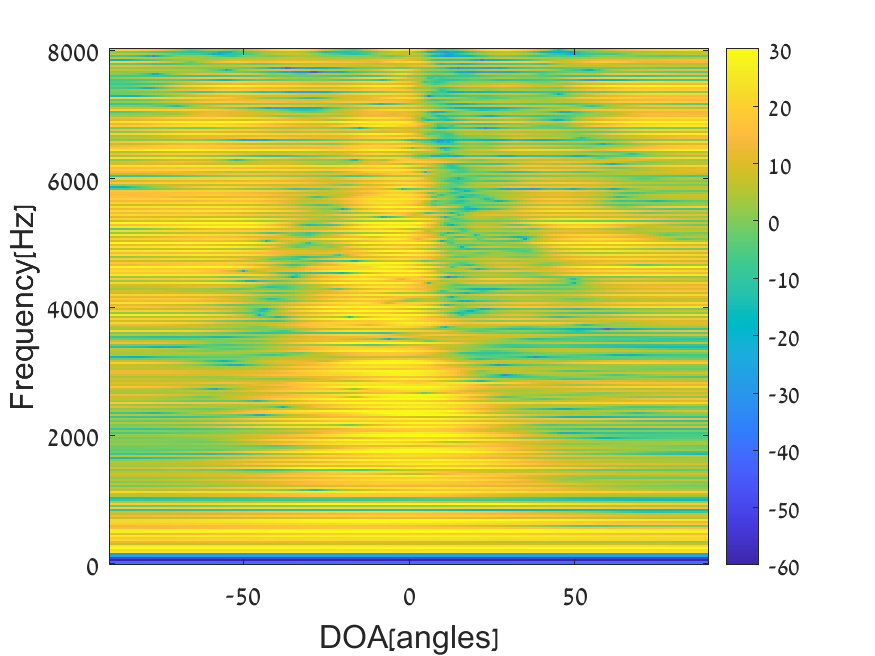}
        \caption{UNET}
        \label{fig:FreqVSDoa_WithoutRevUnet}
    \end{subfigure}\hfill
    \begin{subfigure}{0.45\columnwidth}
        \includegraphics[width=\textwidth]{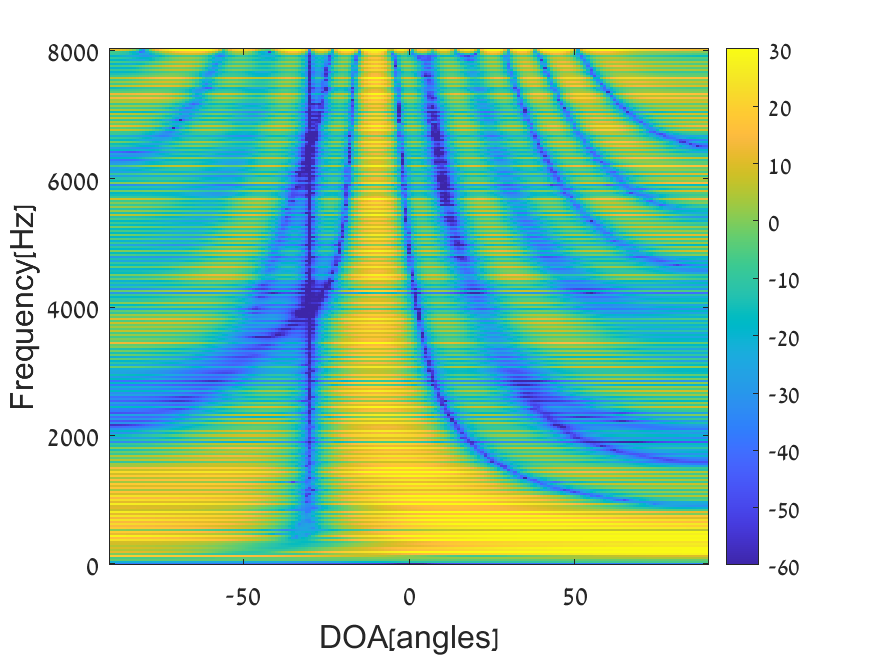}
        \caption{MVDR}
        \label{fig:FreqVSDoa_WithoutRevMVDR}
    \end{subfigure}   
    \caption{$B(k,\theta)$ comparison for a dataset in a non-reverberant environment, where the source is located at $350^\circ$, and the directional noise is at $330^\circ$.}
    \label{fig:beampatternWithoutRevFreq}
\end{figure}

\begin{figure}[t]
    \centering
    \begin{subfigure}{0.45\columnwidth}
        \includegraphics[width=\textwidth]{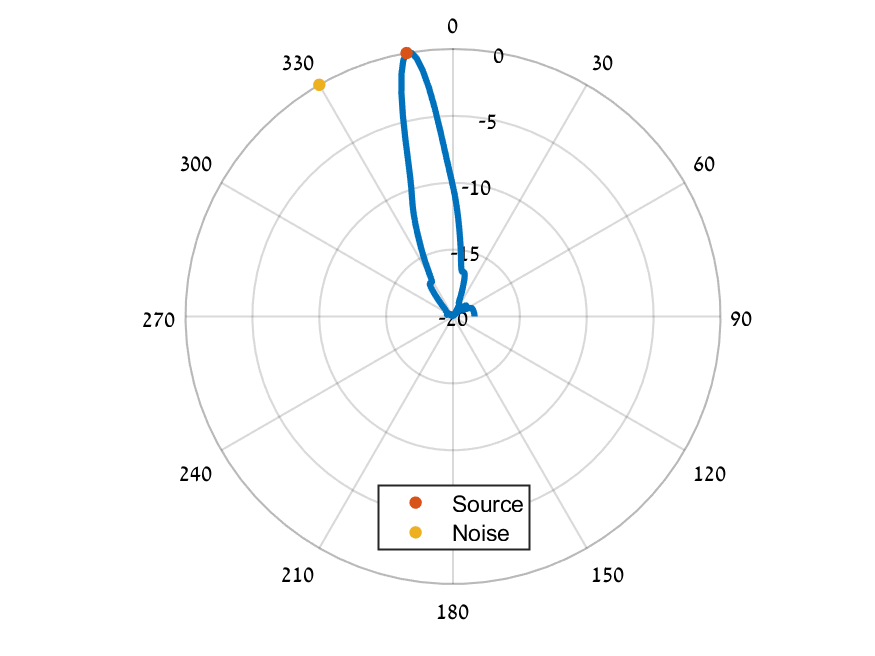}
        \caption{ExNet-BF+PF}
        \label{fig:WithoutRevProposed}
    \end{subfigure}\hfill
    \begin{subfigure}{0.45\columnwidth}
        \includegraphics[width=\textwidth]{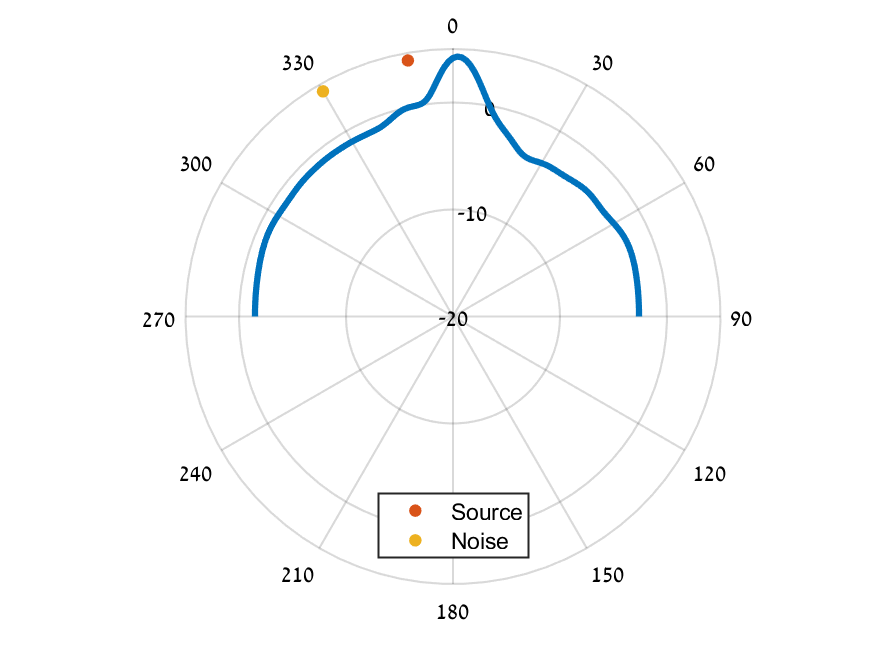}
        \caption{CUNET}
        \label{fig:WithoutRevCUNET}
    \end{subfigure}   
    \begin{subfigure}{0.45\columnwidth}
        \includegraphics[width=\textwidth]{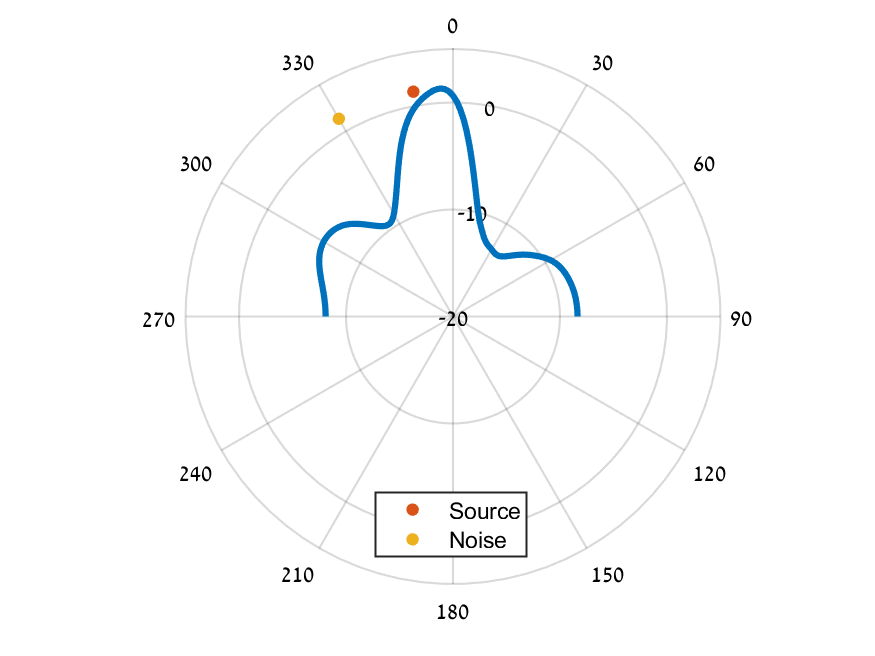}
        \caption{UNET}
        \label{fig:WithoutRevUnet}
    \end{subfigure}\hfill
    \begin{subfigure}{0.45\columnwidth}
        \includegraphics[width=\textwidth]{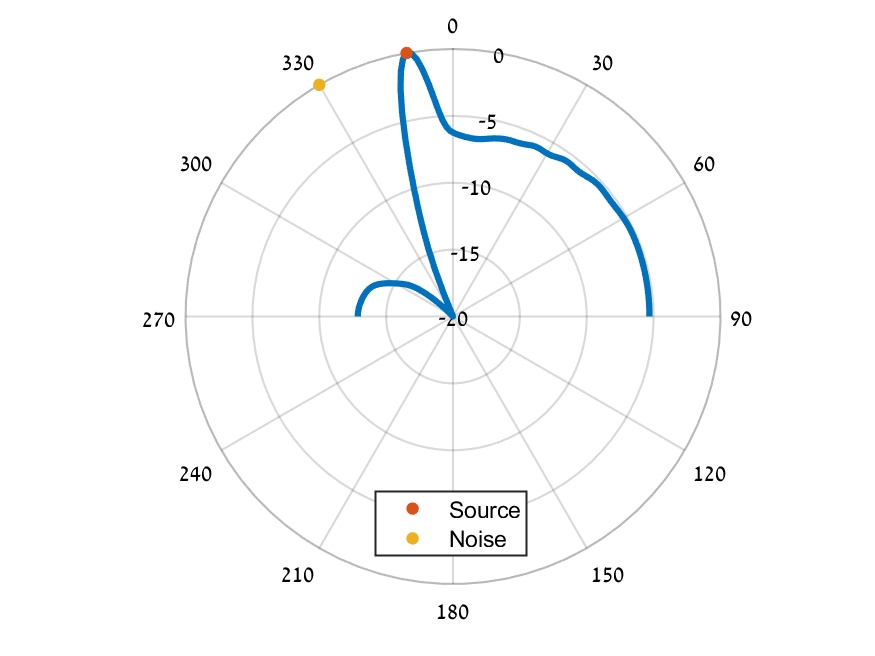}
        \caption{MVDR}
        \label{fig:WithoutRevMVDR}
    \end{subfigure}
    
    \caption{$P(\theta)$ comparison for a dataset in a non-reverberant environment, where the source is located at $350^\circ$, and the directional noise is at $330^\circ$.}
    \label{fig:beampatternWithoutRev}
\end{figure}

For the analysis, we generated an example from the test dataset where the source \ac{DOA} is $350^\circ$, and the directional noise \ac{DOA} is $330^\circ$. The narrow-band beampatterns $|B(k,\theta)|$ for the non-reverberant dataset are depicted in Fig.~\ref{fig:beampatternWithoutRevFreq}, and the corresponding wide-band beampower $P(\theta)$ in Fig.~\ref{fig:beampatternWithoutRev}. The figures clearly indicate that ExNet-BF+PF (we only analyze the first stage output) and the traditional \ac{MVDR} method can direct a beam toward the desired speaker. The CUNET and UNET beamformers do not direct a beam in the correct direction; in any case, the main lobe is not well-pronounced. Comparing the proposed method with the \ac{MVDR} beamformer shows that the main beam of the ExNet-BF+PF is narrower. However, it fails to direct a null toward the directional noise. 

\subsubsection{Dataset in a reverberant environment} 
\begin{figure}[t]
    \centering
    \begin{subfigure}{0.45\columnwidth}
        \includegraphics[width=\textwidth]{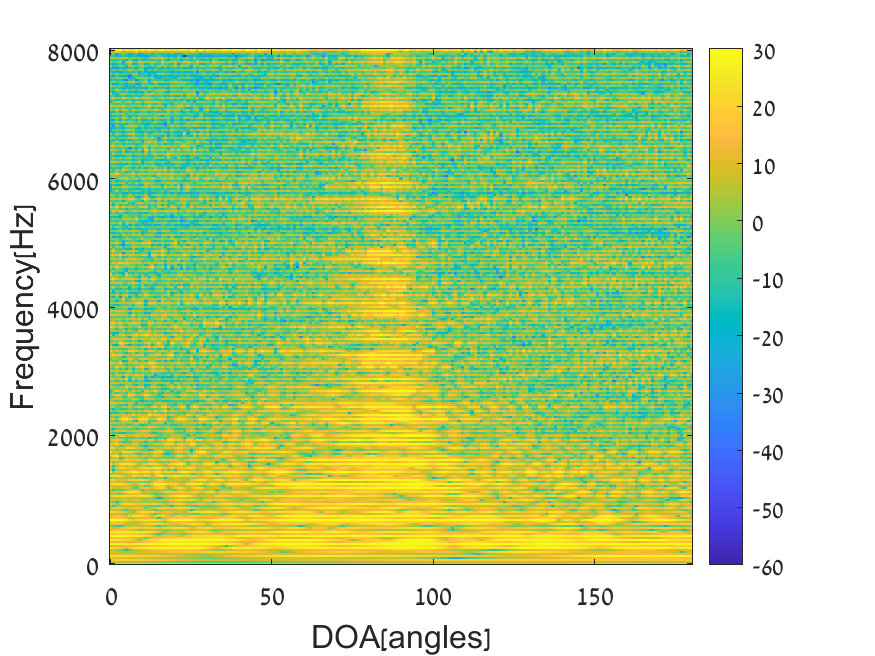}
        \caption{ExNet-BF+PF}
        \label{fig:FreqVSDoa_WithRevProposed}
    \end{subfigure}\hfill
    \begin{subfigure}{0.45\columnwidth}
        \includegraphics[width=\textwidth]{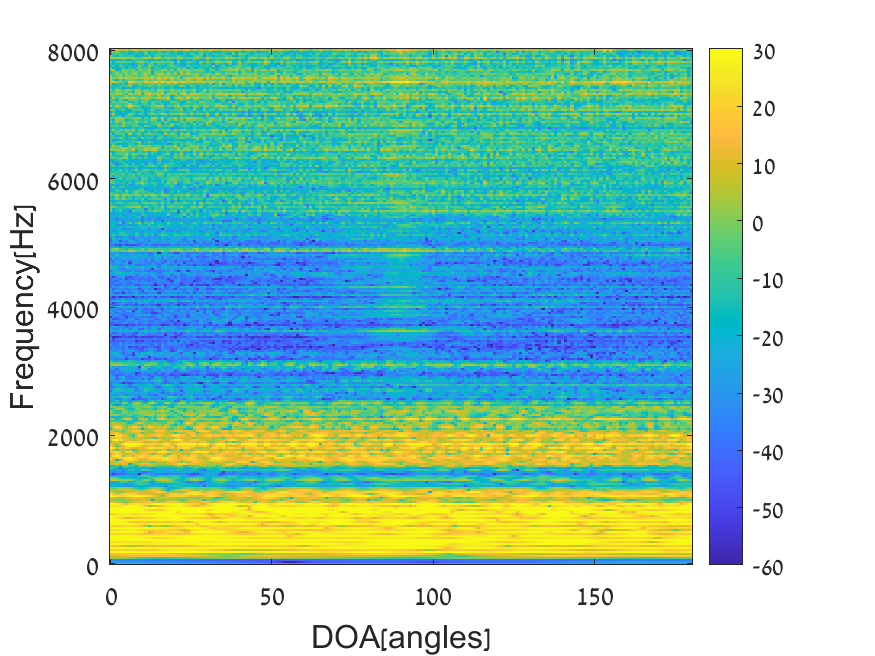}
        \caption{CUNET}
        \label{fig:FreqVSDoa_WithRevCUNET}
    \end{subfigure}    
    \begin{subfigure}{0.45\columnwidth}
        \includegraphics[width=\textwidth]{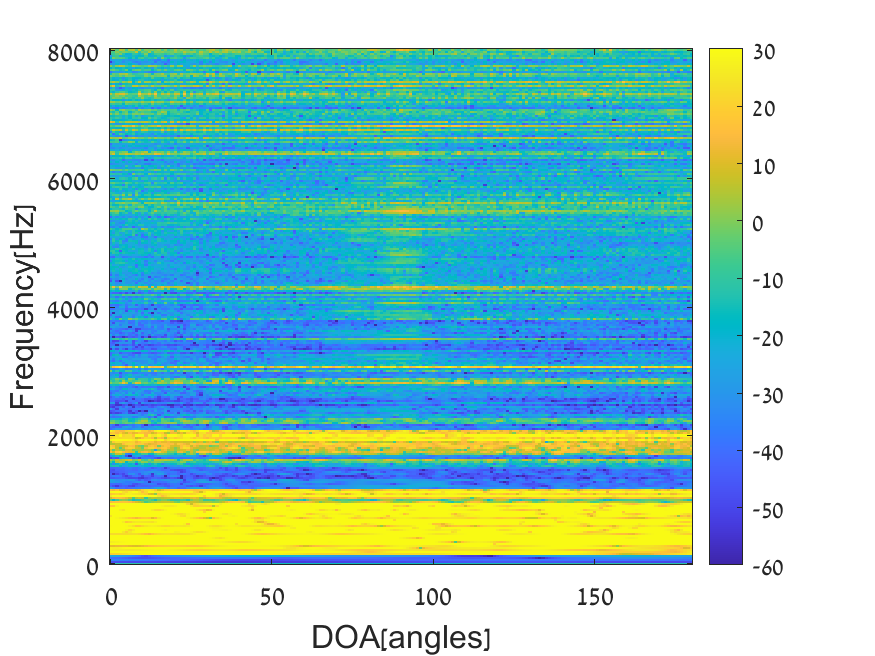}
        \caption{UNET}
        \label{fig:FreqVSDoa_WithRevUnet}
    \end{subfigure}\hfill
    \begin{subfigure}{0.45\columnwidth}
        \includegraphics[width=\textwidth]{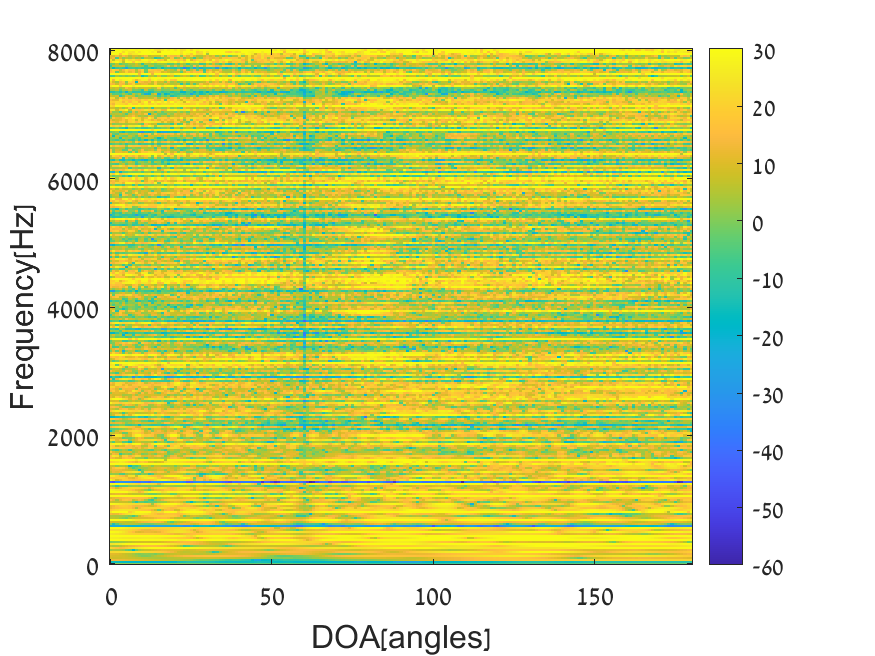}
        \caption{MVDR}
        \label{fig:FreqVSDoa_WithRevMVDR}
    \end{subfigure}    
    \caption{$B(k,\theta)$ comparison for a dataset in a reverberant environment, where the source is located at $350^\circ$, and the directional noise is at $330^\circ$. The reverberation time $T_{60} = 0.4$ seconds.}
    \label{fig:beampatternWithRevFreq}
\end{figure}

\begin{figure}[t]
    \centering
    \begin{subfigure}{0.45\columnwidth}
        \includegraphics[width=\textwidth]{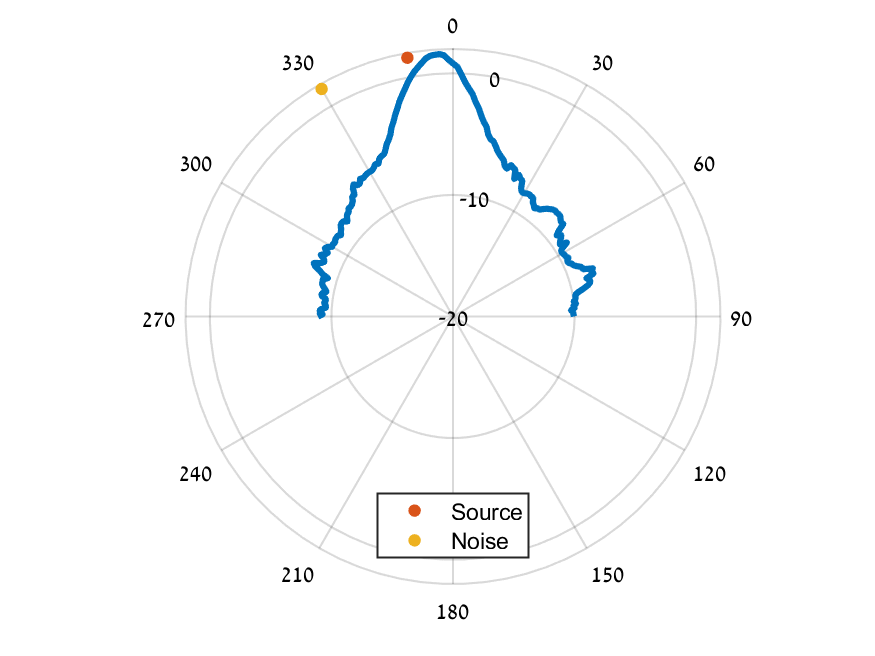}
        \caption{ExNet-BF+PF}
        \label{fig:WithRevProposed}
    \end{subfigure}\hfill
    \begin{subfigure}{0.45\columnwidth}
        \includegraphics[width=\textwidth]{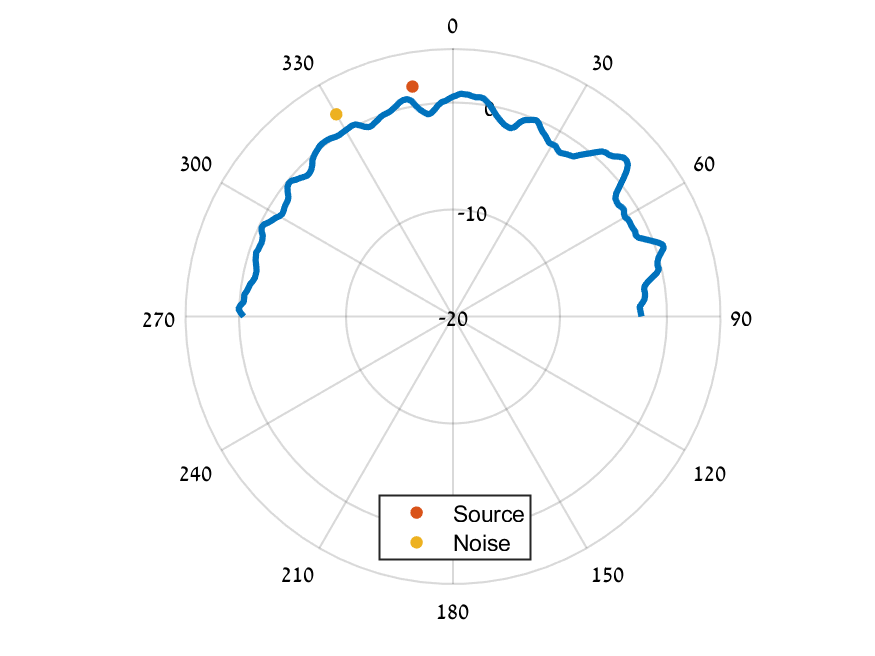}
        \caption{CUNET}
        \label{fig:WithRevCUNET}
    \end{subfigure}    
    \begin{subfigure}{0.45\columnwidth}
        \includegraphics[width=\textwidth]{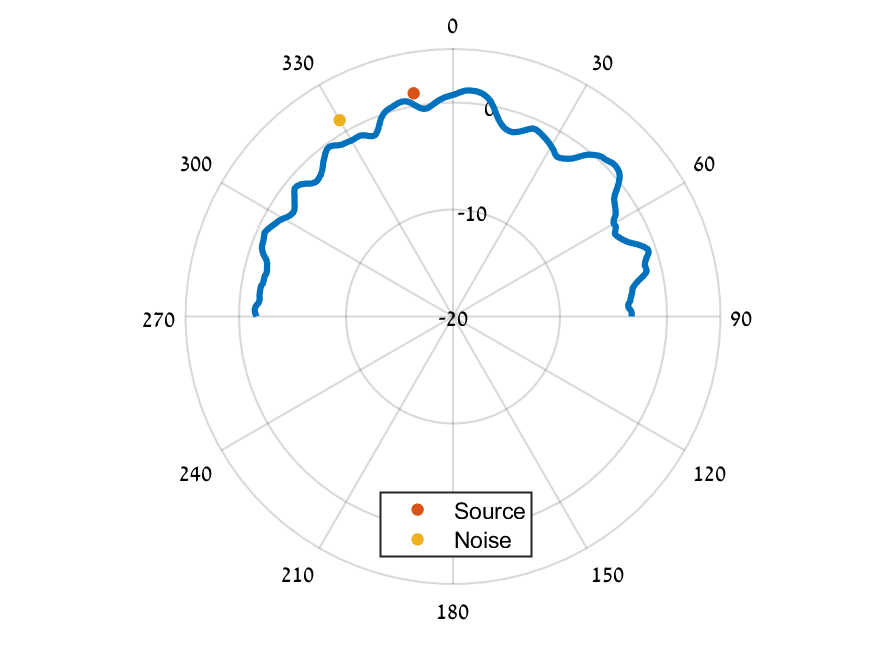}
        \caption{UNET}
        \label{fig:WithRevUnet}
    \end{subfigure}\hfill
    \begin{subfigure}{0.45\columnwidth}
        \includegraphics[width=\textwidth]{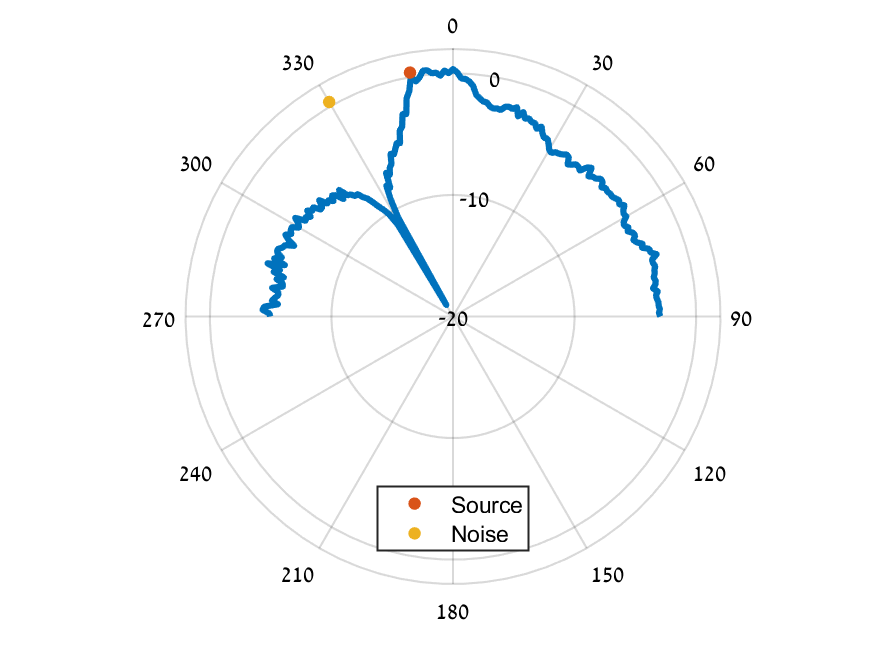}
        \caption{MVDR}
        \label{fig:WithRevMVDR}
    \end{subfigure}    
    \caption{$P(\theta)$, comparison for a dataset in a reverberant environment, where the source is located at $350^\circ$, and the directional noise is at $330^\circ$. The reverberation time $T_{60} = 0.4$ seconds.}
    \label{fig:beampatternWithRev}
\end{figure}

In the case of a dataset in a reverberant environment, obtaining a beampattern proved challenging compared to the results observed in a non-reverberant environment. We adopted a two-step training approach to address this issue and successfully generate a beampattern for a reverberant environment. The first stage, namely the network's multi-channel processor, which implements a time-invariant beamformer, was frozen with weights derived from training on a non-reverberant dataset, while in the second stage, the single-channel postfilter, weights obtained from training in a reverberant environment were utilized.
For this analysis, we generated an example from the test dataset with the same source and directional noise positions as in the previous analysis but with different room properties. Figure.~\ref{fig:beampatternWithRevFreq} depicts the results for $|B(k,\theta)|$ for the reverberant environment dataset, accompanied by $P(\theta)$ results in Fig.~\ref{fig:beampatternWithRev}. Similar to the non-reverberant environment dataset, our proposed approach and the traditional \ac{MVDR} method are the only beamformers capable of creating a directional beam aimed at the speaker. Despite challenges in directing a null towards the directional noise and directing a beam precisely to the speaker's angle, our proposed method shows a clearly defined main lobe compared to the \ac{MVDR}, as depicted in Fig.~\ref{fig:beampatternWithRevFreq}.

In forthcoming discussions of the results on datasets with reverberation, we attribute the success of the ExNet-BF+PF to the two-step training approach manifested by its capability to generate a beampattern.

\subsubsection{Investigating the absence of a null in the beampattern} 
Examining the beampattern of ExNet-BF+PF, it is clear that the proposed beamformer does not produce a null toward the directional interference, in contrast to the \ac{MVDR} beamformer. We experimented with various approaches to impose a null. 

First, we tried to incorporate additional regularization term in the loss function related to the directional noise. Specifically, we add the term $\textrm{mean}(|n_d(t)|)$ to the loss function, where $n_d(t)$ corresponds to $x_d(t)$ in \eqref{eq:loss_mae}, applying on the directional noise rather than the speech signal. Additionally, we trained the network on a dataset where room dimensions and microphone array location were randomized while preserving fixed angles of the speaker and directional noise with respect to the array axis. However, these efforts yielded no success, indicating that the current implementation of our network fails to generate a null effectively.

\subsection{Results}
Audio test samples and their corresponding spectrograms can be found on the project page\footnote{Project page: \href{https://exnet-bf-pf.github.io/}{https://exnet-bf-pf.github.io/}}.
\subsubsection{Non-reverberant environment} 
The results for the noisy dataset are presented in Table \ref{table:withoutRevResults}. First, our network demonstrates superior performance in the \ac{NR} metric. Additionally, it outperforms the other \ac{DNN} models in all evaluation measures. However, in direct comparison with the conventional \ac{MVDR} approach, the \ac{MVDR} beamformer yields better results. It is essential to emphasize that the \ac{MVDR} method relies on prior identification of segments exclusively containing noise, a condition that does not apply to the \ac{DNN} models. Furthermore, the two-stage decomposition of the network contributes to performance enhancement, as discussed in the previous section, emphasizing its additional advantages in spatial analysis.

\subsubsection{Reverberant environment} 
A performance comparison between the proposed algorithm and the competing approaches is summarized in Table~\ref{table:withRevResults}. In this case, our network outperforms the \ac{MVDR} in terms of \ac{NR}, \ac{SISDR}, and \ac{PESQ} metrics, although the \ac{MVDR} still leads in \ac{STOI} and \ac{ESTOI}. Compared with the other \ac{DNN} models, our network leads in all measures except for the \ac{SISDR} metric, where the SC method takes the lead. It is crucial to acknowledge that the performance drop in \ac{SISDR} performance results from our two-step training approach. Albeit this performance loss, this procedure achieves our aim of preserving the network's spatial structure. Given our method's superior performance in other metrics, we opted for this approach despite the slight degradation in \ac{SISDR} performance.

\begin{table}[htbp]
\caption{Results for a dataset in a non-reverberant environment}
\begin{center}
\resizebox{\columnwidth}{!}{
\begin{tabular}{@{}lccccc@{}}
    \toprule
    Input                   & 75.01 & 49.83 & 3.00 & 1.06 & 5.22 
    \\
    \midrule
    Model                   & $\Delta$STOI & $\Delta$ESTOI &  $\Delta$SISDR & $\Delta$PESQ & $\Delta$NR\\
    \midrule
    ExNet-BF+PF             & 18.84 & 34.78 & 12.20 & 1.10 & \textbf{54.45} \\
    SC                      & 10.46 & 17.28 & 7.82  & 0.43 & 41.35 \\ 
    CUNET~\cite{CausalUnet} & -0.60 & -1.27 & 3.43  & 0.15 & 27.68 \\
    UNET                    & 0.98  & -1.88 & 5.93  & 0.16 & 32.21 \\ 
    MVDR+PF                 & \textbf{24.10} & \textbf{47.80} & \textbf{18.94} & \textbf{2.52} & 41.43 \\
    \bottomrule
\end{tabular}}
\end{center}
\label{table:withoutRevResults}
\end{table}

\begin{table}[htbp]
\caption{Results for a dataset in a reverberant environment}
\begin{center}
\resizebox{\columnwidth}{!}{
\begin{tabular}{@{}lccccc@{}}
    \toprule
    Input                   & 66.35 & 47.55 & 2.98 & 1.13 & 5.37 \\
    \midrule
    Model                   & $\Delta$STOI & $\Delta$ESTOI &  $\Delta$SISDR & $\Delta$PESQ & $\Delta$NR\\
    \midrule
    ExNet-BF+PF             & 15.60 & 17.17 & 3.80 & \textbf{0.73} & \textbf{49.86} \\
    SC                      & 13.05 & 13.05 & \textbf{6.46} & 0.61 & 43.64 \\ 
    CUNET~\cite{CausalUnet} & 3.70 & -0.05 & 3.68 & 0.07 & 40.38  \\
    UNET                    & 3.50 & -1.06 & 3.48 & 0.01 & 43.44 \\ 
    MVDR+PF                 & \textbf{16.76} & \textbf{20.55} & 2.04 & 0.59 & 20.74 \\
    \bottomrule
\end{tabular}}
\end{center}
\label{table:withRevResults}
\end{table}

\subsubsection{Noise Types} 
\begin{figure}[ht]
    \centering
    \begin{subfigure}{0.45\textwidth}
        \includegraphics[width=\textwidth]{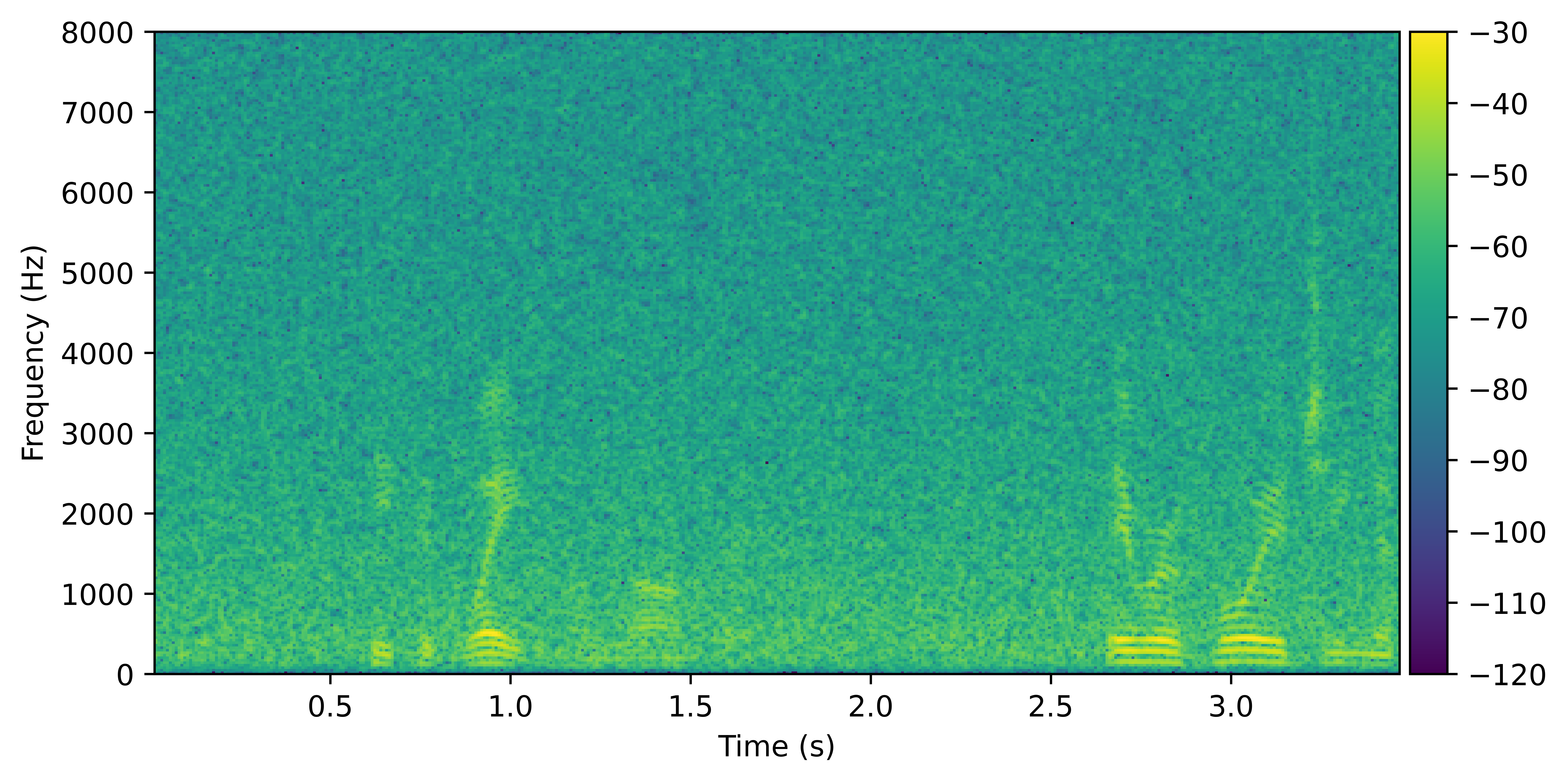}
        \caption{Noisy Signal}
        \label{fig:Time_Varying_Noise_Input}
    \end{subfigure}\hfill
    \begin{subfigure}{0.45\textwidth}
        \includegraphics[width=\textwidth]{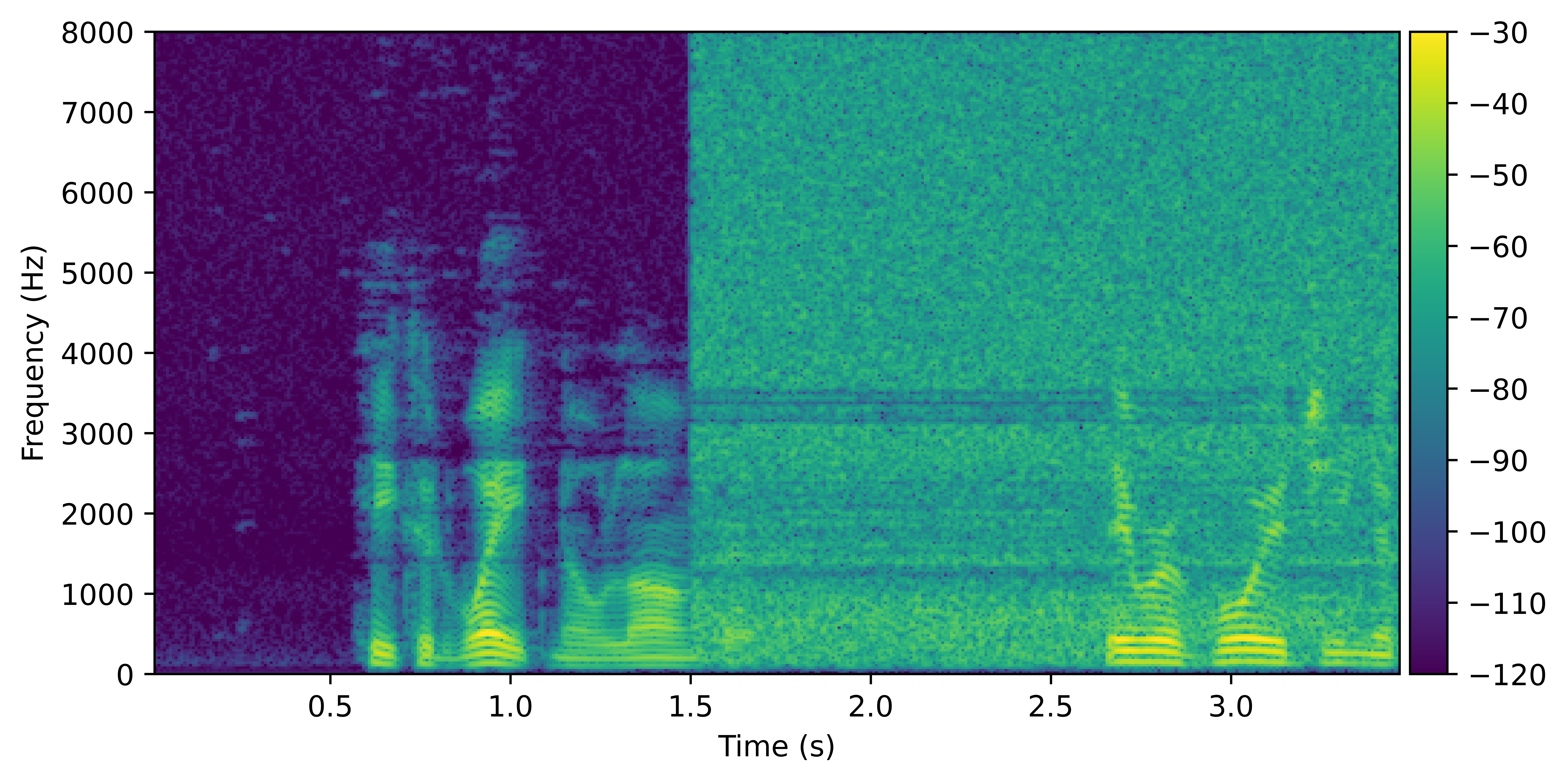}
        \caption{MVDR}
        \label{fig:Time_Varying_Noise_MVDR}
    \end{subfigure}\hfill  
    \begin{subfigure}{0.45\textwidth}
        \includegraphics[width=\textwidth]{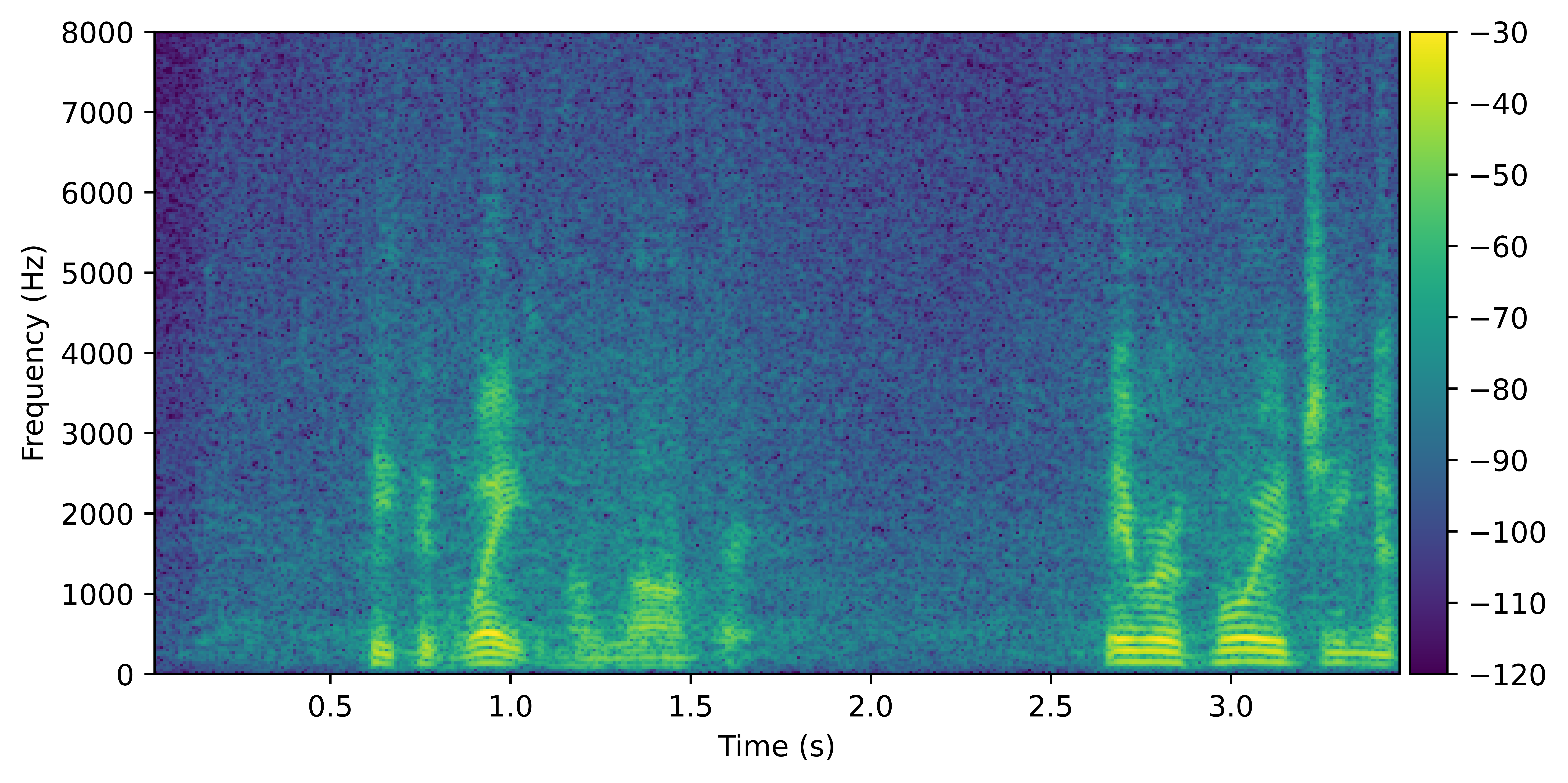}
        \caption{SC}
        \label{fig:Time_Varying_Noise_SC}
    \end{subfigure} \hfill 
    \begin{subfigure}{0.45\textwidth}
        \includegraphics[width=\textwidth]{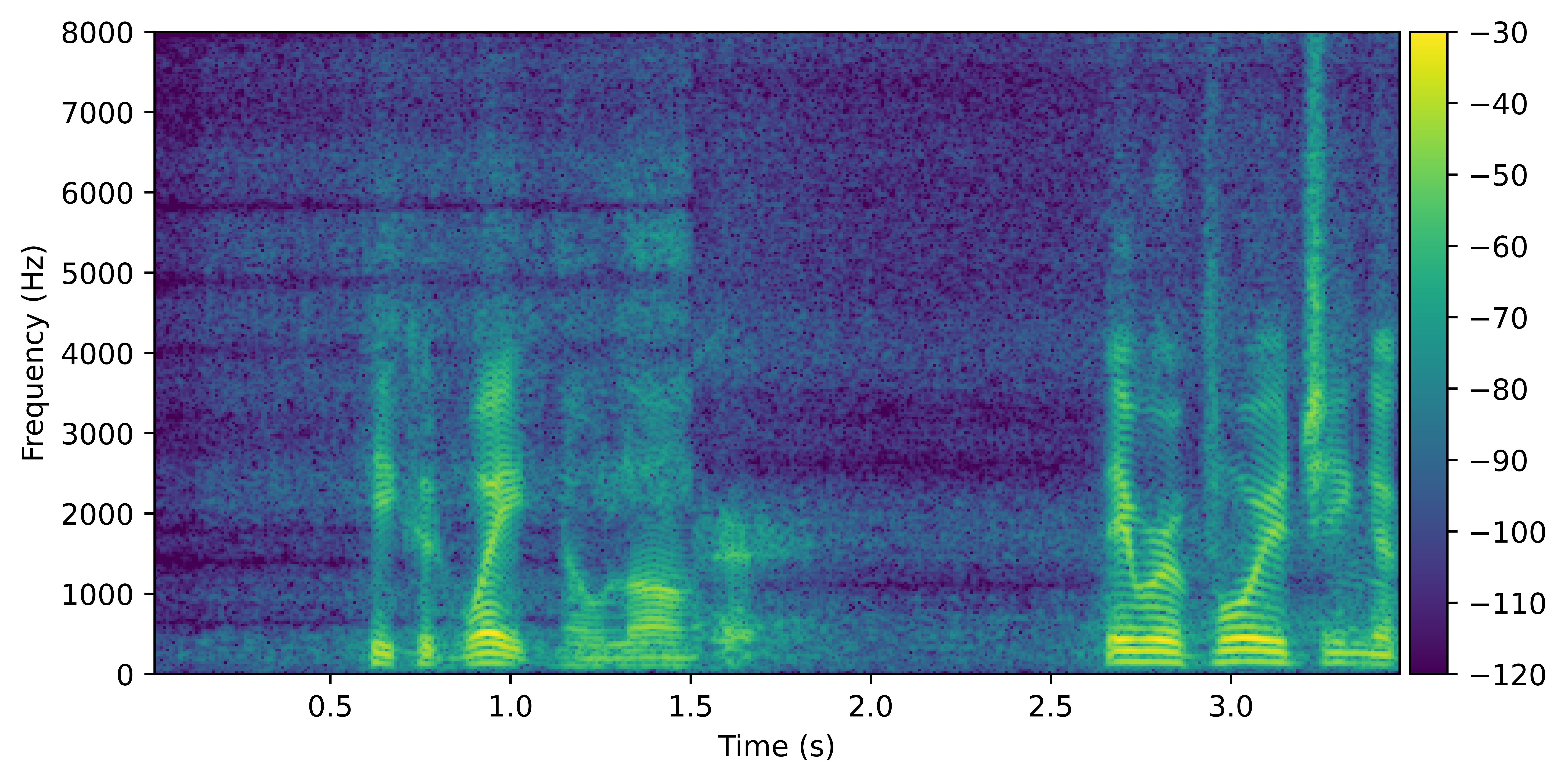}
        \caption{ExNet-BF+PF}
        \label{fig:Time_Varying_Noise_Net}
    \end{subfigure} 
    \caption{Spectrograms showing the Time-Varying Noise scenario in a non-reverberant dataset.}
    \label{fig:Time_Varying_Noise_Fig}
\end{figure}

\paragraph{Non-Reverberant Environment}
The performance of all competing algorithms for different noise types is presented in Table~\ref{table:specialCases} for non-reverberant conditions. For Time-Varying Noise and Babble Voice scenarios, our method outperforms competing approaches. Conversely, the \ac{MVDR} method maintains its lead in most metrics in scenarios involving Babble Noise and Speaker Switch. Notably, our method performs well without relying on prior knowledge, showcasing its robustness compared to the \ac{MVDR} method. Additionally, our network consistently outperforms the other \ac{DNN}-based methods in all scenarios. Figure~\ref{fig:Time_Varying_Noise_Fig} depicts the spectrograms of an example from the Time-Varying Noise case. These results clearly illustrate the challenges faced by \ac{MVDR}, particularly evident in the latter part of the signal, where the noise statistics have changed. Moreover, our network excels in noise reduction compared to both \ac{MVDR} and SC methods.

\begin{figure}[ht]
    \centering
    \begin{subfigure}{0.45\textwidth}
        \includegraphics[width=\textwidth]{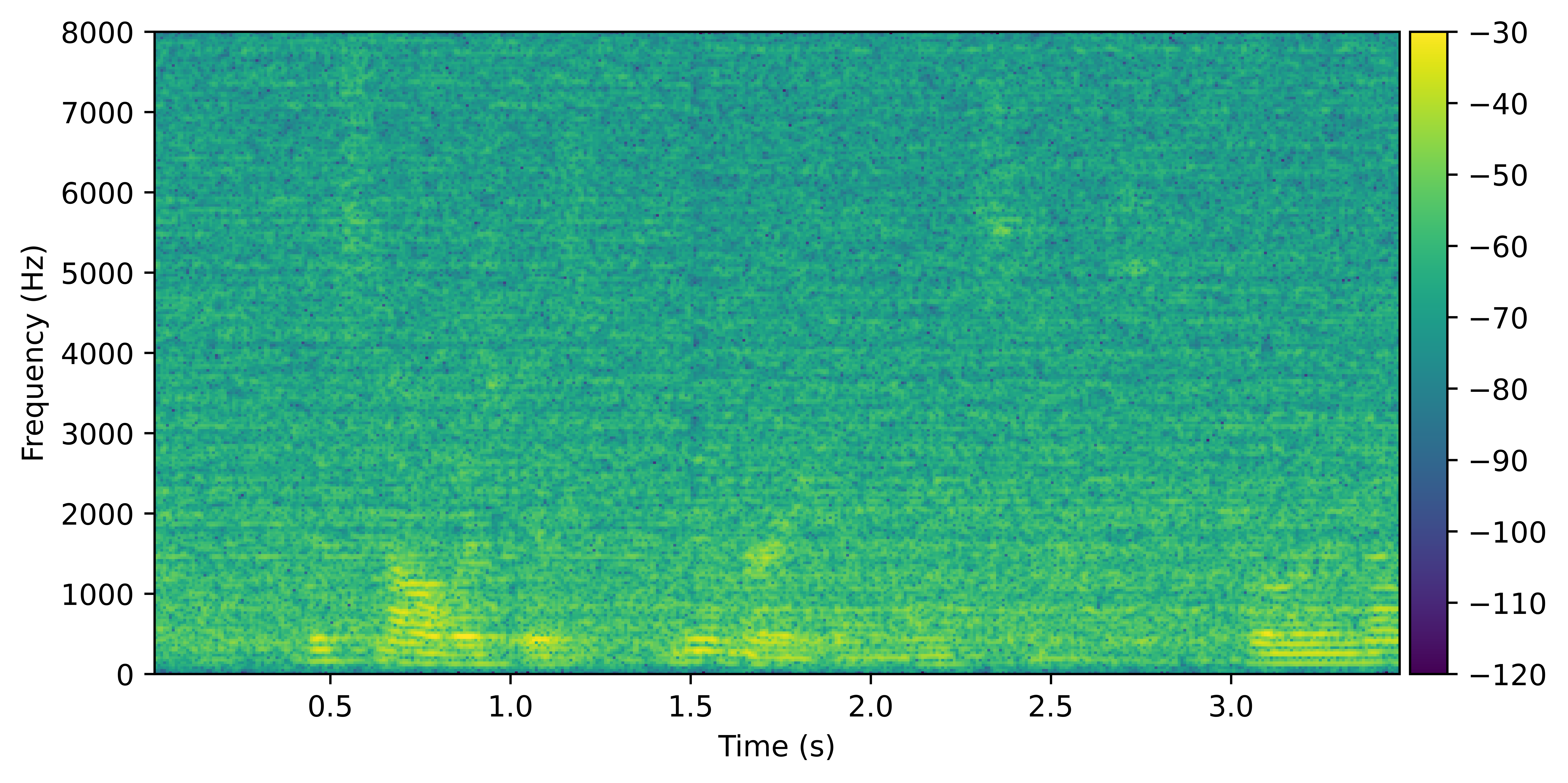}
        \caption{Noisy Signal}
        \label{fig:Time_Varying_Noise_Input_withRev}
    \end{subfigure}\hfill
    \begin{subfigure}{0.45\textwidth}
        \includegraphics[width=\textwidth]{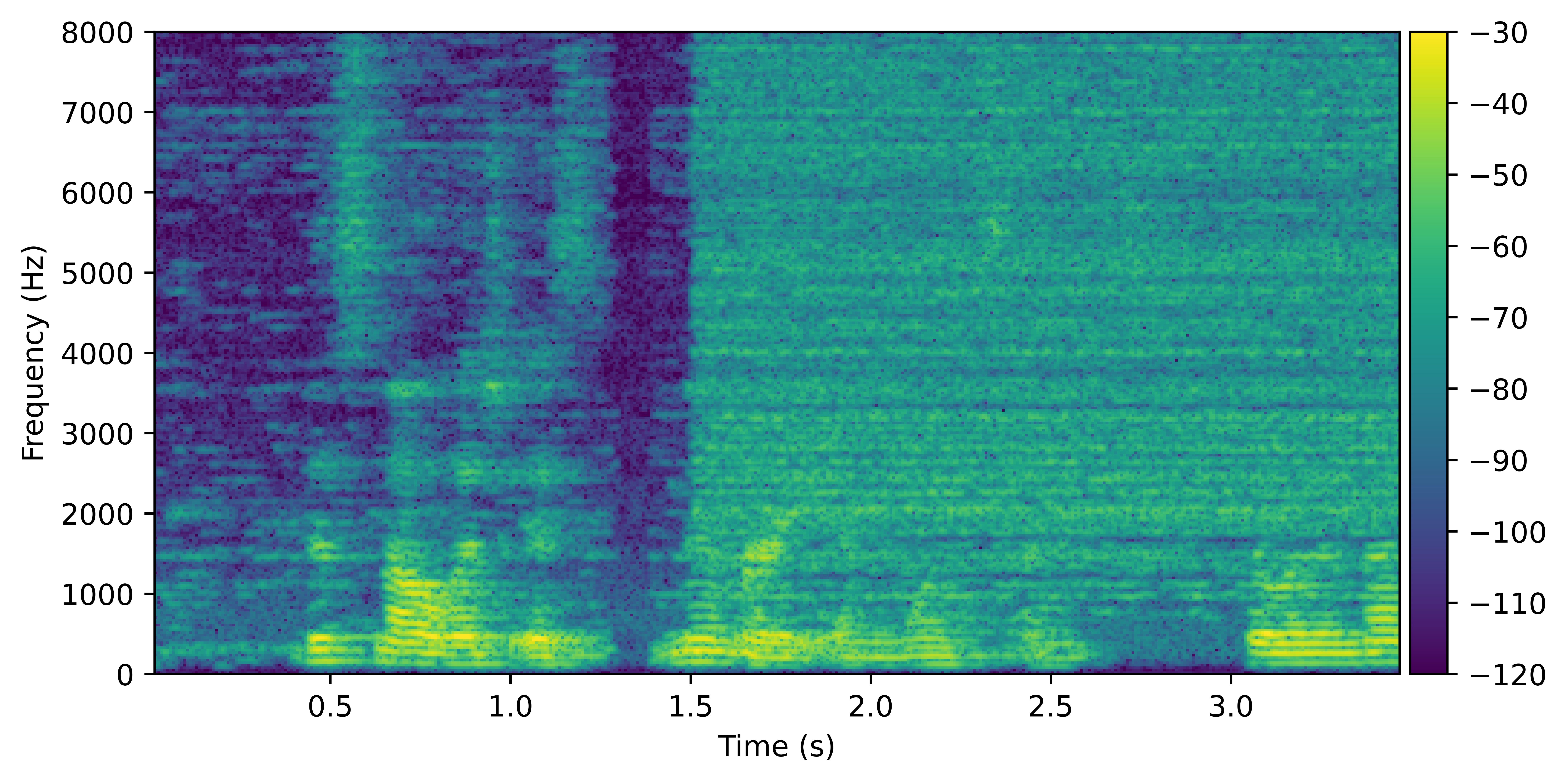}
        \caption{MVDR}
        \label{fig:Time_Varying_Noise_MVDR_withRev}
    \end{subfigure}\hfill  
    \begin{subfigure}{0.45\textwidth}
        \includegraphics[width=\textwidth]{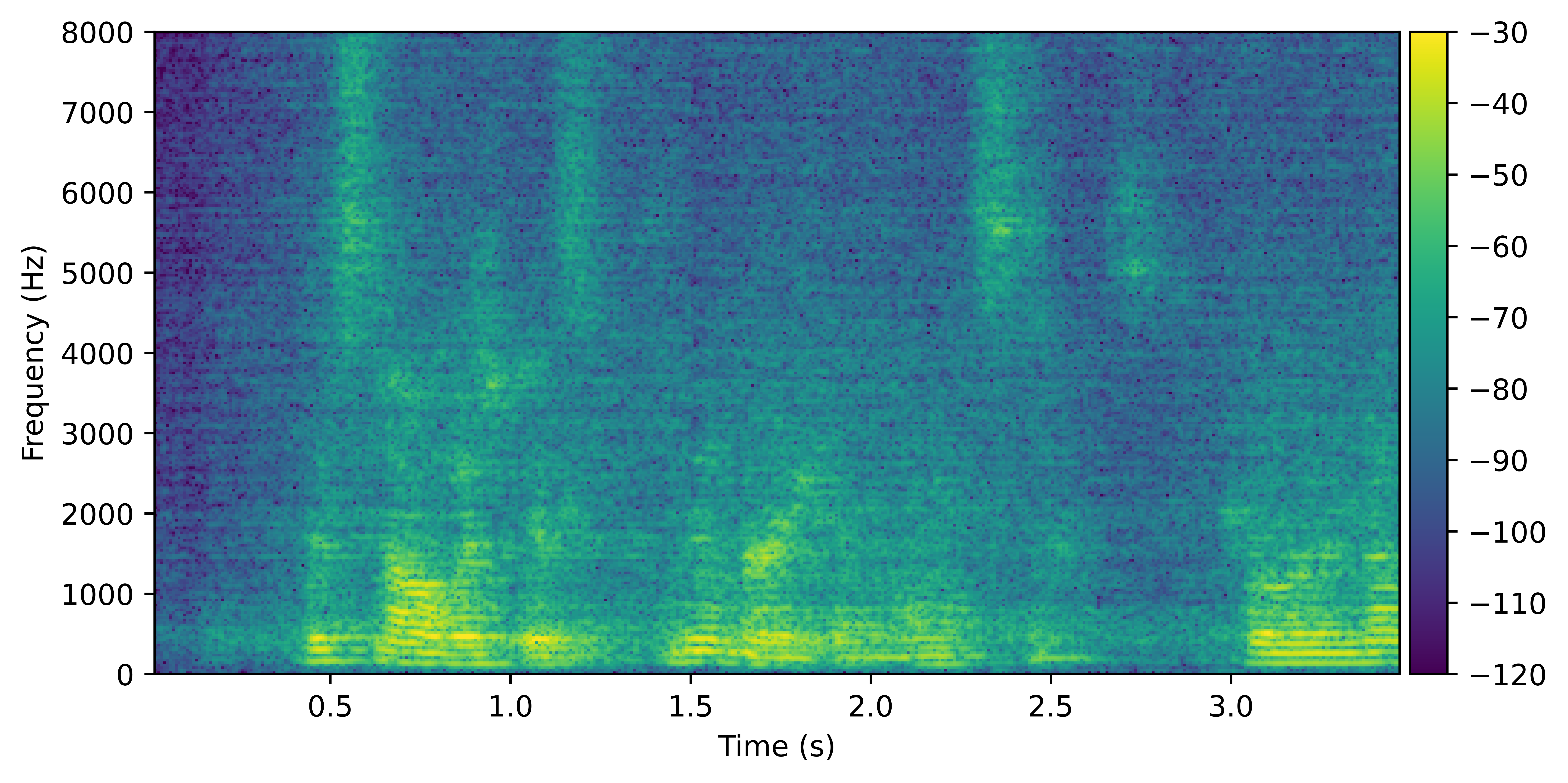}
        \caption{SC}
        \label{fig:Time_Varying_Noise_SC_withRev}
    \end{subfigure} \hfill 
    \begin{subfigure}{0.45\textwidth}
        \includegraphics[width=\textwidth]{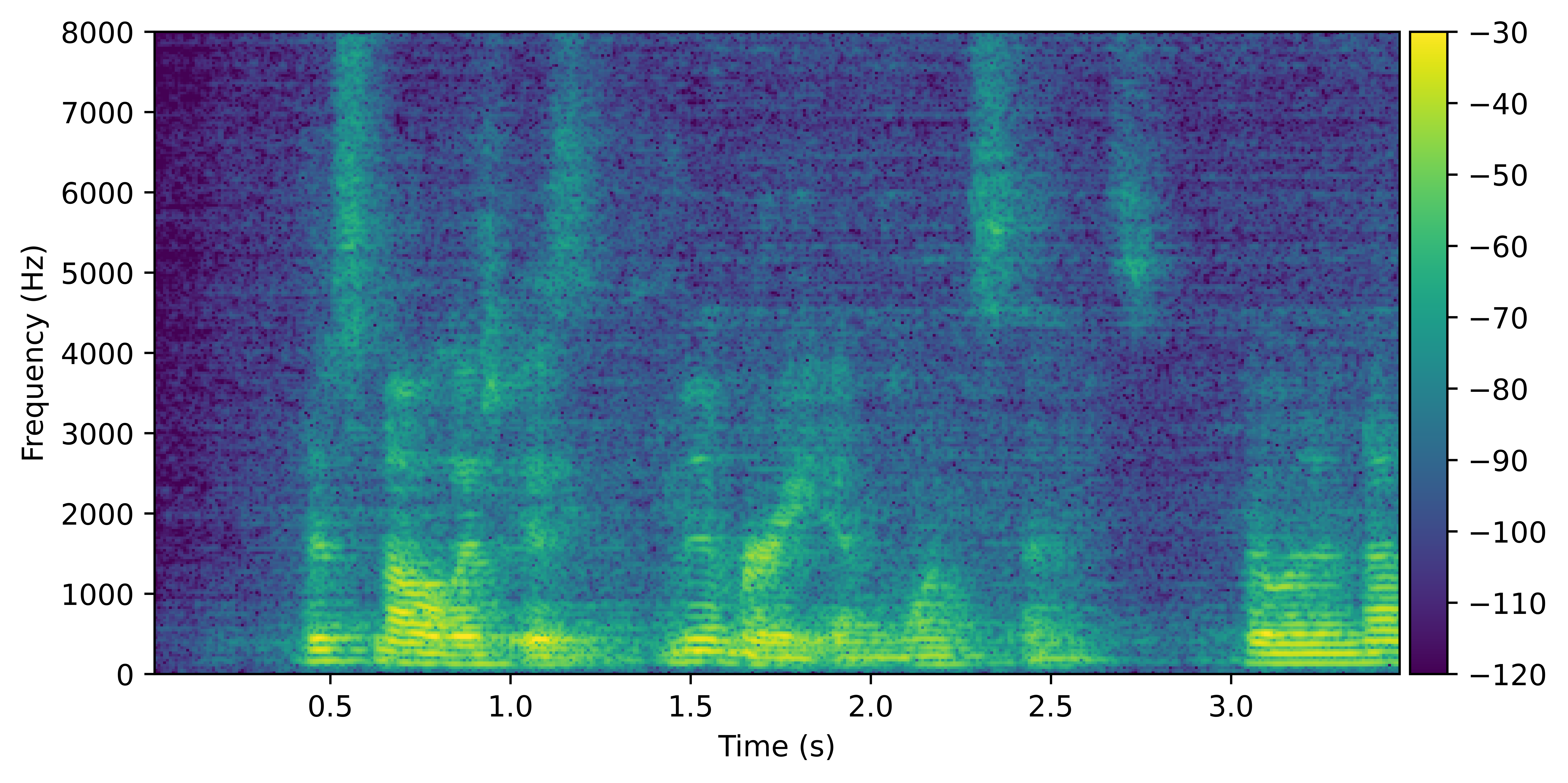}
        \caption{ExNet-BF+PF}
        \label{fig:Time_Varying_Noise_Net_withRev}
    \end{subfigure} 
    \caption{Spectrograms showing the Time-Varying Noise scenario in a reverberant dataset.}
    \label{fig:Time_Varying_Noise_Fig_withRev}
\end{figure}

\paragraph{Reverberant Environment}
Examining results with a reverberant dataset, as presented in Table~\ref{table:specialCasesWithRev}, further emphasizes the strength of our network. Even in scenarios involving Babble Noise and Speaker Switch in the presence of reverberation, our network surpasses the performance of \ac{MVDR}. While SC did not outperform in any metric in the non-reverberant scenario, it consistently yields better \ac{SISDR} results across all cases in the reverberant dataset, as mentioned earlier. Figure~\ref{fig:Time_Varying_Noise_Fig_withRev} presents the spectrograms of an example from the
Time-Varying Noise case. These samples clarify that the \ac{MVDR} approach cannot track the noise statics switch. Additionally, as observed in non-reverberant environments, our neural network demonstrates superior noise reduction performance compared to both \ac{MVDR} and SC methods.

\begin{table*}[htbp]
\caption{Results for various noise types in a non-reverberant environment.}
\begin{center}
\resizebox{1.2\columnwidth}{!}{
\begin{tabular}{@{}llccccc@{}}
\toprule
Noise Type & Model & $\Delta$STOI & $\Delta$ESTOI &  $\Delta$SISDR & $\Delta$PESQ & $\Delta$NR\\
\midrule
\multirow{6}{*}{Time-Varying Noise} & Input & 74.11 & 48.86 & 2.98 & 1.07 & 5.17 \\
\cmidrule(l){2-7}
& ExNet-BF+PF & \textbf{19.28} & \textbf{35.06} & \textbf{12.05} & \textbf{1.12} &  \textbf{54.39} \\
& SC & 10.85 & 17.66 & 7.82 & 0.48 & 41.43 \\
& CUNET~\cite{CausalUnet} & -0.73 & -1.14 & 3.35 & 0.20 & 27.40 \\
& UNET & 1.53 & -1.65 & 5.96 & 0.27 & 31.97 \\
& MVDR+PF & 5.05 & 13.47 & 2.19 & 0.21 & 44.67 \\ 
\midrule
\multirow{6}{*}{Speaker Switch} & Input & 75.40 & 50.48 & 3.00 & 1.06 & 5.20 \\
\cmidrule(l){2-7}
& ExNet-BF+PF & 14.71 & 25.85 & \textbf{8.74} & 0.69 & \textbf{50.96} \\
& SC & 10.68 & 17.86 & 8.24 & 0.45 & 41.58 \\
& CUNET ~\cite{CausalUnet} & -0.03 & 0.10 & 3.84 & 0.20 & 27.97 \\
& UNET & 2.79 & 0.90 & 6.42 & 0.24 & 32.47 \\
& MVDR+PF & \textbf{20.88} & \textbf{42.24} & 0.06 & \textbf{1.31} & 40.35 \\ 
\midrule
\multirow{6}{*}{Babble Noise} & Input & 73.54 & 48.94 & 3.00 & 1.05 & 5.23 \\
\cmidrule(l){2-7}
& ExNet-BF+PF & 18.98 & 32.75 & 11.74 & 0.86 & \textbf{54.44} \\
& SC & 11.21 & 17.05 & 7.75  & 0.42 & 41.59 \\
& CUNET~\cite{CausalUnet} & 0.84 & 0.70 & 3.57 & 0.17 & 30.56 \\
& UNET & 3.92 & 2.50 & 6.10 & 0.22 & 37.18 \\
& MVDR+PF & \textbf{20.51} & \textbf{37.89} & \textbf{13.77} & \textbf{1.33} & 26.36 \\ 
\midrule
\multirow{6}{*}{Babble Voice} & Input & 78.84 & 56.05 & 2.98 & 1.18 & 9.36 \\
\cmidrule(l){2-7}
& ExNet-BF+PF & \textbf{12.00} & \textbf{22.56} & \textbf{7.82} & \textbf{0.49} & \textbf{52.98} \\
& SC & 5.53  & 11.13 & 5.41 & 0.22 & 43.74 \\
& CUNET~\cite{CausalUnet} & -11.54 & -10.30 & -0.49 & -0.08 & 41.16 \\
& UNET & -10.93 & -12.52 & 0.21 & -0.10 & 48.79 \\
& MVDR+PF & 10.73 & 19.33 & 6.37 & 0.32 & 12.40 \\
\bottomrule
\end{tabular}}
\end{center}
\label{table:specialCases}
\end{table*}

\begin{table*}[htbp]
\caption{Results for various noise types in a reverberant environment.}
\begin{center}
\resizebox{1.2\columnwidth}{!}{
\begin{tabular}{@{}llccccc@{}}
\toprule
Noise Type & Model & $\Delta$STOI & $\Delta$ESTOI &  $\Delta$SISDR & $\Delta$PESQ & $\Delta$NR\\
\midrule
\multirow{6}{*}{Time-Varying Noise} & Input & 66.57 & 47.52 & 2.98 & 1.12 & 5.33 \\
\cmidrule(l){2-7}
& ExNet-BF+PF & \textbf{15.17} & \textbf{16.72} & 3.61 & \textbf{0.69} & \textbf{49.63} \\
& SC & 12.66 & 12.52 & \textbf{6.28} & 0.57 & 43.49 \\
& CUNET~\cite{CausalUnet} & 2.83 & -0.64 & 3.23 & 0.09 & 40.23 \\
& UNET & 2.71 & -2.02 & 3.00 & 0.02 & 43.16 \\
& MVDR+PF & 0.67 & 1.97 & -0.26 & 0.14 & 21.30 \\
\midrule
\multirow{6}{*}{Speaker Switch} & Input & 68.92 & 49.46 & 2.99 & 1.09 & 5.36 \\
\cmidrule(l){2-7}
& ExNet-BF+PF & \textbf{12.93} & \textbf{14.84} & 4.99 & \textbf{0.58} & \textbf{47.12} \\
& SC & 11.89 & 12.92 & \textbf{6.66} & 0.55 & 43.84 \\
& CUNET ~\cite{CausalUnet} & 2.22 & -0.35 & 3.68 & 0.15 & 40.46 \\
& UNET & 2.26 & -1.35 & 3.55 & 0.08 & 43.34 \\
& MVDR+PF & 10.27 & 14.40 & -1.88 & 0.50 & 19.69 \\
\midrule
\multirow{6}{*}{Babble Noise} & Input & 67.62 & 48.12 & 2.99 & 1.13 & 5.43 \\
\cmidrule(l){2-7}
& ExNet-BF+PF & \textbf{15.06} & \textbf{16.71} & 3.84 & \textbf{0.64} & \textbf{50.40} \\
& SC & 12.85 & 12.89 & \textbf{6.30} & 0.53 & 44.07 \\
& CUNET~\cite{CausalUnet} & 2.36 & -1.25 & 3.12 & 0.01 & 40.01 \\
& UNET & 2.73 & -1.83 & 3.06 & -0.02 & 42.53 \\
& MVDR+PF & 10.22 & 11.24 & 2.15 & 0.41 & 18.60 \\
\midrule
\multirow{6}{*}{Babble Voice} & Input & 71.47 & 54.56 & 2.99 & 1.24 & 10.32 \\
\cmidrule(l){2-7}
& ExNet-BF+PF & \textbf{6.49} & 6.08 & 1.41 & \textbf{0.27} & \textbf{51.39} \\
& SC & 6.19 & \textbf{6.23} & \textbf{4.04} & 0.22 & 47.35 \\
& CUNET~\cite{CausalUnet} & -7.71 & -10.99 & -0.86 & -0.15 & 35.93 \\
& UNET & -9.02 & -11.70 & -1.19 & -0.17 & 38.56 \\
& MVDR+PF & 0.73 & -0.56 & -1.30 & 0.07 & 7.22 \\
\bottomrule
\end{tabular}}
\end{center}
\label{table:specialCasesWithRev}
\end{table*}

\section{Conclusions}
\label{sec:conclusion}

This article introduced an explainable DNN-based beamformer with a postfilter. Our method leverages a U-Net network architecture combined with a beamformer structure. In our implementation, an attention mechanism is incorporated into the skip connections of the U-Net architecture. Our approach offers distinct advantages by decomposing the network into spatial and spectral components through a two-stage processing scheme. This decomposition facilitates analysis and explainability of the beamforming architecture, together with performance advantages in most cases. The results demonstrate consistent positive outcomes across all metrics, although sometimes closely matching other methods. Overall, our method performs well without requiring prior knowledge of the speaker's activities, particularly in scenarios involving various noise types.
Furthermore, we conducted a comprehensive analysis of the spatial characteristics of our network as compared with competing approaches. Our method effectively utilizes spatial information to enhance the signal, evidenced by its ability to direct a focused beam toward the speaker. Although the current architecture cannot direct a null toward a directional noise, it still significantly improves signal quality through advanced spatial information utilization.

\bibliographystyle{IEEEtran}
\bibliography{my_library}

\end{document}